\documentclass[usegraphicx,usenatbib]{mn2e}
\usepackage{times}
\usepackage{epsf}

\newcommand{\source}{\hbox{2152-699}}
\newcommand{\sourcefull}{\hbox{PKS\,B2152-699}}

\newcommand{\chandra}{\textit{Chandra}}

\newcommand{\hst}{\textit{HST}}

\newcommand{\atca}{\textit{ATCA}}
\newcommand{\wfi}{\textit{WFI}}

\title[\sourcefull: new deep radio and X-ray observations]
{The jet-cloud interacting radio galaxy
\sourcefull. I. Structures revealed in new deep radio and X-ray observations}

\author[D.M. Worrall et al.]
 {D.M. Worrall$^1$,
M. Birkinshaw$^1$,
A.J.~Young$^1$,
K.~Momtahan$^1$,
R.A.E.~Fosbury$^2$,
\newauthor
R.~Morganti$^{3,5}$,
C.N.~Tadhunter$^4$
and G.~Verdoes Kleijn$^5$
\\
$^1$HH Wills Physics Laboratory, University of Bristol, Tyndall Avenue,
Bristol BS8~1TL \\
$^2$European Southern Observatory, D-85748 Garching bei M\"unchen, Germany \\
$^3$Netherlands Institute for Radio Astronomy, PO Box 2, 7990 AA
Dwingeloo, The Netherlands \\
$^4$Department of Physics and Astronomy, University of Sheffield,
 Sheffield S3~7RH \\
$^5$Kapteyn Astronomical Institute, University of Groningen, Postbus 800,
9700 AV Groningen, The Netherlands 
 }

\begin{document}

\label{firstpage}

\maketitle

\begin{abstract}

PKS B2152-699, which has radio power characteristic of sources that
dominate radio feedback, is exceptional in showing a wide range of
features associated with radio-galaxy/gas interactions.  We present
new deep radio (ATCA), X-ray (\chandra) and ground-based optical
observations, and test the energetics of the feedback model.  We
report the first high-resolution observations of the radio jet,
finding that the inner jet extends
$\sim$ 8.5 kpc ($10^\circ$ viewing angle) in the direction of an optical emission-line High
Ionization Cloud (HIC) before taking a zig-zag path to a position
offset from the HIC.  Jet synchrotron radiation to X-ray energies is
seen.  The HIC is associated with cool, $kT \sim 0.3$-keV, X-ray gas
of anomalously low metallicity.  On larger scales, the radio galaxy
displays all three X-ray features that together confirm supersonic
expansion of the lobes into the external medium: gas cavities,
inverse-Compton emission showing excess internal lobe pressure, and
high-contrast arms of temperature above the $kT \sim 1$~keV ambient
medium.  The well-formed southern lobe on the counterjet side is
expanding with a Mach number between 2.2 and 3.  The lobe energy
appears to be more gently dissipated in the north.  We
estimate a cavity power $\sim 3 \times 10^{43}$ ergs s$^{-1}$, which
falls well below previously reported correlations with radio power.
The total inferred time-averaged jet power, $\sim 4 \times 10^{44}$
ergs s$^{-1}$, is dominated by the kinetic and thermal energy of
shocked gas, and if used instead would bring the source into better
agreement with the correlations.  The southern hotspot is the more
complex, with a spiral polarization structure.  Its bright peak emits
synchrotron X-rays.  The fainter northern hotspot is particularly
interesting, with X-rays offset in the direction of the incoming jet
by $\sim 1$ arcsec relative to the radio peak.  Here modest
($\delta \sim 6$) relativistic beaming and a steep radio spectrum cause the jet to be
X-ray bright through inverse-Compton scattering before it
decelerates. With such beaming, a modest proton content
or small departure from minimum energy in the jet will
align estimates of the instantaneous and time-averaged jet power.
The hotspots suggest acceleration of electrons to a
maximum energy $\sim 10^{13}$ eV in the jet termination shocks.

\end{abstract}


\section{Introduction}
\label{sec:intro}

The source \sourcefull\ (hereafter \source), is one of the brightest
nearby (luminosity distance 123~Mpc, see below) radio galaxies in the
southern sky \citep{schilizzi}.  It has
attracted attention due to the presence of a High Ionization Cloud
(HIC) \citep{tadhunter87, tadhunter88, alighieri} that lies adjacent to a
radio-bright knot near the base of the northern lobe
\citep{fosbury98}, providing evidence for a strong jet-cloud
interaction.  Moreover, X-ray measurements with {\it Chandra\/} have
discovered a large-scale X-ray-emitting atmosphere containing
cavities, interpreted as having been inflated by the lobes of the
radio source \citep{young}.  \source\ is therefore an example of a
radio source where feedback between the AGN and surrounding plasma can
be well studied.

However, rather than just an example, \source\ has particular
importance in the study of radio-source feedback since it is a nearby
example at the characteristic radio power of sources that must
dominate such feedback in the Universe as a whole \citep{wrev}.  The
total radio flux density is 68 Jy at 468 MHz, and $\alpha_{\rm r} \sim
0.8$ \citep[$S_\nu \propto \nu^{-\alpha_{\rm r}}$,][]{kuhr}, giving a
power extrapolated to 178~MHz of $2 \times 10^{25}$ W Hz$^{-1}$
sr$^{-1}$, and placing it in the range where FRIs and FRIIs overlap in
luminosity \citep{fr}.  Its hotspots are embedded in its N and S lobes
rather than at the extremities \citep{fosbury98}, which is a common
morphology for sources close to the FRI/FRII border \citep*{capetti}.
The luminosity function of radio sources places those at the FRI/FRII
boundary as the population that contributes the greatest fraction of
the overall radio output of active galaxies \citep{ledlow, kaiser}.
In combination with correlations between radio power and cavity power
\citep{cavagnolo}, it is therefore apparent that sources at the
FRI/FRII boundary are those of most interest for studies of radio-mode
feedback.  A typical such source is not in a cluster atmosphere,
unlike those to have received most attention to date
\citep[e.g.,][]{forman, gitti, fabian}.  Characteristically, there is
nothing extraordinary about the environment of \source, increasing its
importance as an exemplar of radio-source feedback.

Aside from the large-scale properties of \source, study of the
jet and hotspots has been limited to date by a lack of sensitive
high-resolution radio mapping extending from the core to hotspots, rectified in this paper.
Earlier papers have demonstrated that the jet must undergo
deflection in order to terminate in the N hotspot.  However, where
that deflection occurs has remained uncertain.  The VLBI measurements
of \citet{tingay} found a pc-scale nuclear radio jet at position angle
$44\pm4$ deg from the core, whereas the N hotspot is at an angle of
about 24 deg from the core.  This led the authors to suggest that the
jet is deflected by about 20 deg due to oblique shocks within the HIC,
which also lies at a position angle of about 44 deg from the core.
The situation became more complicated with the detection of radio
emission from a knot lying close to the HIC but sufficiently separated
to be at a position angle of about 34 deg from the core
\citep{fosbury98}.  The 18-GHz radio map presented in this paper
reveals where the jet bends, and in conjunction with multiwavelength
data we find some clues as to why.

Section 2 describes our new \atca, \chandra, and ground-based optical
observations, and our new processing of \hst\ data. Measurements from
these data are presented in Section 3, and in Section 4 we describe
separately our deductions concerning the northern radio jet, the two
hotspots, the HIC, and feedback as inferred from the properties of the
radio lobes and hot-gas environment. Conclusions are summarized in
Section 5.  We adopt $H_0 = 70$~km~s$^{-1}$~Mpc$^{-1}$, $\Omega_{m_0}
= 0.3$, $\Omega_{\Lambda_0} = 0.7$.  1~arcsec corresponds to 566~pc at
the redshift ($z=0.0282$) of \source, and the luminosity distance is
$D_{\rm L} = 123$~Mpc. Spectral index, $\alpha$, is defined in the
sense that flux density is proportional to $\nu^{-\alpha}$: the photon
spectral index measured in X-ray observations is $1 + \alpha$, and the
energy index of the number spectrum of synchrotron-radiating electrons
is $2\alpha + 1$.

\begin{figure}
\centering
\includegraphics[width=3.2truein]{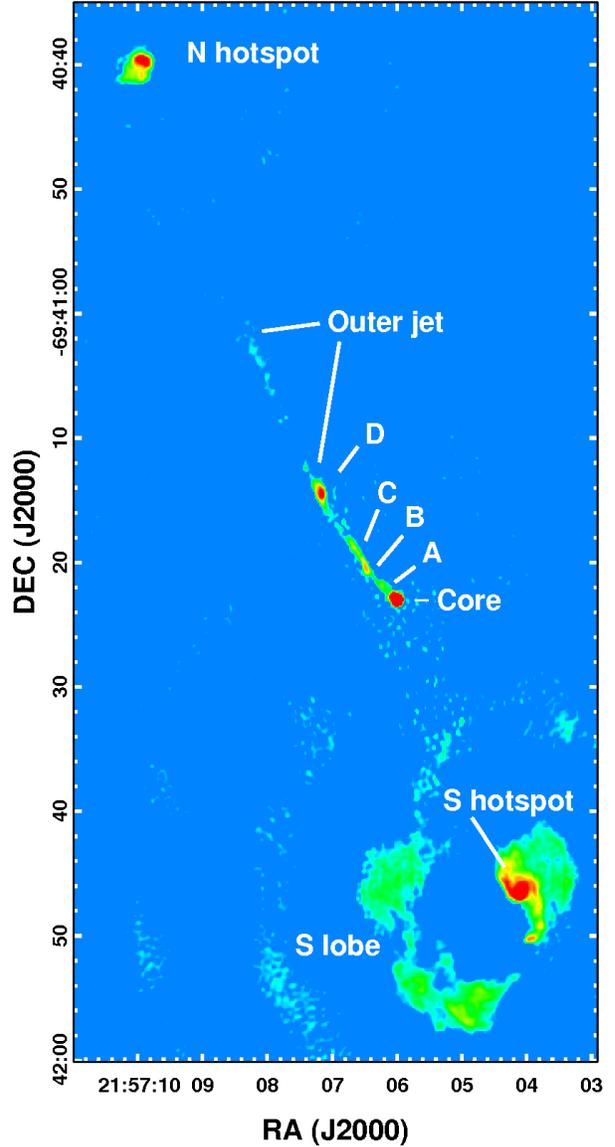}
\caption{
18-GHz \atca\ radio map with 0.35 arcsec restoring beam and source
features labelled.  The faint features to the east of the S lobe are
residual grating rings from the bright S hotspot.}
\label{fig:mapfull}
\end{figure}

\section{Observations and analysis methods}
\label{sec:obs}%

\subsection{\atca\ radio}
\label{sec:obsR}%

We made a full-track observation of \source\ with the \atca\ in its 6A
array configuration with the Compact Array Broadband Backend (CABB)
correlator \citep{wilson} on 2009 June 22nd (programme C2034).  The
observations were in two frequency bands centered at 17.0 and 19.0
GHz, each of bandwidth 2048 MHz divided into 2048 channels.  The
sources B1921-293 and B1934-638 were observed for bandpass and primary
flux calibration, respectively.  B2353-686, adjacent to \source, was
observed for phase calibration and as a pointing reference.  Data
reduction was carried out with the {\sc miriad}
\footnote{http://www.atnf.csiro.au/computing/software/miriad} software
using the calibration advice for CABB data given in the 2010 September
version of the Miriad User Guide.  

Large time-dependent phase slopes were seen across the frequency bands,
presumably due to atmospheric delays.  To help correct for this,
each band was divided into eight separate sub-bands in frequency for
individual self-calibration and multi-frequency synthesis cleaning,
and results were combined for final mapping.  The final map
(Fig.~\ref{fig:mapfull}) was made with uniform weighting and a 0.35
arcsec restoring beam.  Noise in the map was dominated by residual
phase errors. Nevertheless, except very close to the bright core, the
dynamic range near the jet is 24,000:1, which is a notable achievement
for a complex source at such a high frequency with a single \atca\
east-west array configuration.  It is of historical interest that
\source\ was the first source to be imaged with the \atca\ during
construction in 1989, at 5~GHz with a 30 arcsec FWHM beam
\citep{norris}.

We make use of a map with a 2.35-arcsec restoring beam made from
pre-CABB \atca\ 4.74~GHz data taken in 1992 and presented in
\citet{fosbury98}.  We applied an astrometric correction of 1.06
arcsec, mainly in declination, to align the core as seen on that map
with that in our 18~GHz map.

\subsection{\chandra\ X-ray}
\label{sec:obsX}%

We made a 125~ks observation of \source\ in full window and FAINT data
mode with the Advanced CCD Imaging Spectrometer (ACIS) on board
\chandra\ in 2010.  The observation was divided into exposures on
January 20th (OBSID 12088) and January 22nd/23rd (OBSID 11530).  The
source was positioned at the normal aimpoint of the front-illuminated
I3 chip. The other three CCDs of ACIS-I and the S2 and S3 chips of
ACIS-S were also on during the observations, giving a frame time of
3.24~s.  Details of the instrument and its modes of operation can be
found in the \chandra\ Proposers' Observatory
Guide\footnote{http://cxc.harvard.edu/proposer}.  Results presented
here use {\sc ciao v4.3} and the {\sc caldb v4.4.5} calibration
database.  We re-calibrated the data to take advantage of the
sub-pixel event reposition routine ({\sc edser}), following the
software ``threads'' from the \chandra\ X-ray Center
(CXC)\footnote{http://cxc.harvard.edu/ciao}, to make new level~2
events files.  Only events with grades 0,2,3,4,6 were retained. After
a small loss of exposure time due to eliminating periods where the
background rate was more than 3$\sigma$ from the average, the
calibrated datasets have durations of 58.359~ks (OBSID 12088) and
56.352~ks (OBSID 11530).

A shorter earlier observation in full window and VFAINT data mode had
been made with \chandra\ on 2001 August 2nd, as described by
\citet*{ly} and \citet{young}. For that observation \source\ was
positioned at the normal aimpoint of the back-illuminated S3 chip, and
the S1, S2, S4, I2 and I3 chips were also on.  We re-calibrated those
data, as above, and applied VFAINT cleaning to help remove particle
background.  That dataset has a final exposure time of time of
12.192~ks (OBSID 1627).

The astrometry of the X-ray images was adjusted to align with the
18-GHz radio core.  Shifts were roughly 0.75 arcsec, mainly in
declination.  Our analysis uses combined data from the three
observations.  For display purposes we have made a merged
exposure-corrected file.  For spectral analysis we have fitted the
same model parameters jointly to the individual data-sets, using their
respective calibration files.

For spectral fitting we have used {\sc
xspec}\footnote{http://heasarc.nasa.gov/xanadu/xspec}.  Where features
are too weak to allow sufficient spectral bins containing at least 20
counts we have estimated parameter confidence ranges using the CSTAT
statistic applied to on-source and background regions separately. We
used the APEC model for thermal components.  All fits include
absorption along the line of sight in our Galaxy assuming a column
density of $N_{\rm H} = 2.52 \times 10^{20}$ cm$^{-2}$ [from the {\sc
colden} program provided by the CXC, using data of \citet{dlock90}].
Quoted errors are 1$\sigma$ taking into account the number of
interesting parameters.

\subsection{\wfi\ optical}
\label{sec:wfi}

We observed \source\ with the MPG/ESO 2.2-m 34-arcmin-field-of-view
Wide Field Imager (\wfi) using a V-band filter (number 843) for 3~ks
in October 2004.  While the primary objective was to search for [O
III]-emitting clouds aligned with the radio axes but further from the
nucleus than the HIC, we use the data here to report optical emission
from \source's N and S hotspots, which lie beyond the field of view of
the \hst\ observation ({Section~\ref{sec:hst}).  The data were
reduced with two independent software packages, {\sc alambic}
\citep{vandame} and {\sc astro-WISE} \citep{mcfarland}, with
fluxes agreeing to within $\sim 10$ per cent and
astrometry to $\sim 0.1$ arcsec.  The photometric calibration yielded
a systematic error in the zero point of $\sim 0.1$ magnitudes when
calibrated to Johnson V. This budget includes errors in the extinction
and colour terms of the standard stars, and varying illumination over
the \wfi\ chips.  Johnson V magnitude was converted to flux density at
$5.5 \times 10^{14}$ Hz using $3.75 \times 10^{-9}$ ergs s$^{-1}$
cm$^{-2}$ \AA$^{-1}$ for zero magnitude. We applied an astrometric
correction of 0.5 arcsec, mainly in declination, to align the optical
nucleus with that in our 18~GHz map.

\subsection{\hst\ optical}
\label{sec:hst}

We make use of an archival \hst\ WFPC2 PC1 image (F606W filter, single
500~s exposure) published by \citet{fosbury98}, that shows well the
structural detail of the HIC.  We further processed the data using the
{\sc iraf ellipse} software to subtract galaxy light from the image,
with a view to unveiling further features associated with the radio
jet and its surroundings.  We applied an astrometric correction of
1.48 arcsec, mainly in right ascension, to align the nucleus with the
18~GHz radio map.  We used the {\sc iraf synphot} package to calibrate
the optical flux densities of features of interest.

\begin{figure}
\centering
\includegraphics[width=3.2truein]{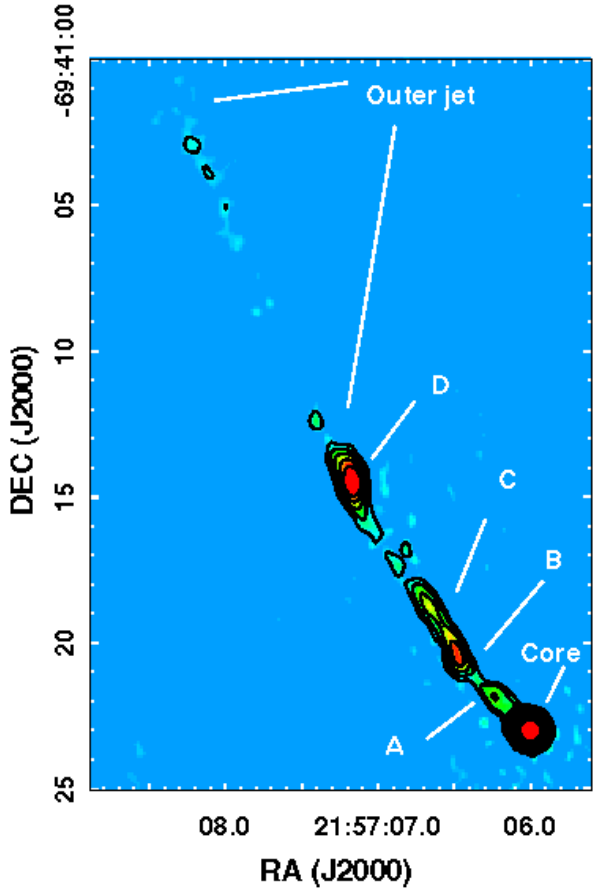}
\caption{
Expanded view of the jet at 18~GHz from the map of Fig.~\ref{fig:mapfull}.  
Contours at $(1,2,4,8,16,64,256,1024) \times$ 0.09 mJy beam$^{-1}$.}
\label{fig:mapjet}
\end{figure}

\begin{figure}
\centering
\includegraphics[width=3.2truein]{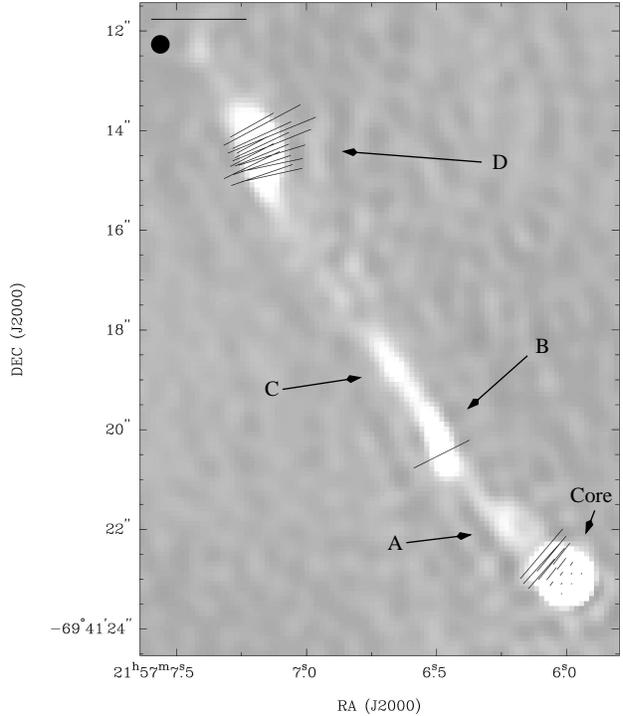}
\caption{
Inner jet shown with its fractional polarization E vectors.  The scale bar at top left
corresponds to 46.5 per cent polarization. The beam size is also shown
at the top left. }
\label{fig:polplot}
\end{figure}

\section{Results}
\label{sec:results}

\subsection{Radio jet}
\label{sec:resultsradio}

Previous radio mapping of \source\ that was carried out with the
\atca\ at 4.7 and 8.6 GHz detected the core, a knot adjacent to the
HIC, and emission from the S and N lobes and their embedded hotspots
\citep{fosbury98}.  In addition, VLBI has measured embedded components
in the core at 8.4 GHz \citep{tingay} and the S hotspot at 1.4 GHz
\citep{young}.  Our new map (Fig.~\ref{fig:mapfull}) is well matched
to the resolution of \chandra\ and is the first to track the jet
through several knots on its way to the N hotspot, and we have
therefore introduced labels for the components.  Knot D lies adjacent
to the HIC.

Intensity and polarization maps for the jet are shown in
Figs.~\ref{fig:mapjet} and \ref{fig:polplot}, respectively, and
discussed in Section~\ref{sec:discussionjet}. All the 18-GHz maps
presented in this paper have a 0.35 arcsec restoring beam.  Results
for the hotspots are presented in Section \ref{sec:resultshotspots}.  and
discussed in Section \ref{sec:discussionhotspots}.

Radio measurements are compiled in Table~\ref{tab:radio}.  For compact
features the parameters are those of elliptical Gaussians fitted using
the {\sc miriad} task {\sc imfit}.  The tabulated polarization
fraction and position angle apply to the centroid of the component in
question, and Figure \ref{fig:polplot}, 
which shows the
E-field vectors, provides an indication of the range of
values over the features.  Radio spectral indices have been estimated
for the brightest components within the 16-20~GHz band of the
observations.  These are rather uncertain, to at least $\pm 0.1$,
since we do not always see a smooth spectrum over our 16 sub-bands of
frequency, presumably due to calibration errors associated with the
time-variable delays present in the data.

\citet{fosbury98} report that between 4.7 and 8.6 GHz the core
spectrum is inverted and may show some historic variability, and our
18-GHz flux density is indeed higher than at 4.7 and 8.6~GHz.  We find
the morphology to be complex, which is not unexpected given that VLBI
finds a core-jet structure with a resolved core \citep{tingay}. We
have fitted the 18-GHz core (Fig.~\ref{fig:mapjet}) to an
unresolved component and a weak jet
component found to be centered 0.35 arcsec away. There is marginal
evidence that the core is slightly elongated in the counter-jet
direction, with a component $\sim 0.1$ arcsec away, although
observations with fuller baseline coverage would be required for
confirmation.

A good fit to knot D requires 3 components. Two of these (D-1 and D-2)
have position angles consistent with the overall jet direction at D,
but the third (D-peak) is inclined by about 10 degrees, in a direction
closer to N-S.  Interestingly, the polarization vector tilts together
with the component orientation, leading to the overall spread in
vectors seen in Fig.~\ref{fig:polplot}.  The 18~GHz spectrum of the
total emission from knot D appears to be rather steep, as noted by
\citet{fosbury98} for lower frequencies.  However,
the D-peak sub-component is distinctly flatter in radio-to-optical spectrum
(Section~\ref{sec:resultsopticaljet}).

\begin{table*}
\caption{Parameters for radio components}
\label{tab:radio}
\begin{tabular}{llllllll}
\hline
(1) & (2) & (3) & (4) & (5) & (6) & (7) & (8)\\
Component & RA, Dec & $S_{\rm 18~GHz}$ (mJy)& FWHM (arcsec) & PA (deg)
&$P\%$ & $\chi$
  (deg) & $\alpha_{\rm r}$\\
\hline
Core & 21:57:$06.000\pm0.0002$,-69:41:$23.001\pm0.001$ & $829\pm15$ &
- & - & 0.6 & $-38$ & -\\
Core-jet & 21:57:$06.049\pm0.002$,-69:41:$22.76\pm0.01$ & $28.3\pm1.2$
& $0.18 \times 0.15$ & - & 20 & $-38$ & -\\
Core-counterjet$\dag$ & 21:57:$05.996\pm0.006$,-69:41:$23.07\pm0.04$ &
$33\pm19$ & $0.18 \times 0.18$ & - & 0.4 & $-38$ & -\\
Core-total & - & $893\pm 16$ & - & - & - & - & 0.15 \\ 
A & 21:57:$06.227\pm0.014$,-69:41:$21.91\pm0.06$ & $3.0\pm0.3$ &
$1.3\times 0.6$ & $49\pm9$ & - & - & -\\
B & 21:57:$06.480\pm0.003$,-69:41:$20.41\pm0.04$ & $4.7\pm0.3$ &
$1.1\times 0.1$ & $20\pm2$ & 30 & $-60$ & -\\
C & 21:57:$06.655\pm0.010$,-69:41:$18.82\pm0.08$ & $3.8\pm0.3$ &
$1.8\times 0.2$ & $33\pm2$ & - & - & -\\
D-1 & 21:57:$07.194\pm0.003$,-69:41:$14.63\pm0.05$ & $5.0\pm0.7$ &
$1.7\times 0.2$ & $24\pm2$ &  30 & $-65$ & -\\
D-peak & 21:57:$07.164\pm0.001$,-69:41:$14.57\pm0.01$ & $5.9\pm0.5$ &
$0.4\times 0.1$ & $8\pm3$ & 30  & $-73$ & -\\
D-2 & 21:57:$07.168\pm0.004$,-69:41:$14.14\pm0.05$ & $4.7\pm0.5$ &
$0.8\times 0.2$ & $26\pm3$ & 30  & $-65$ & -\\
D-total & - & $15.6\pm 1$ & - & - & - & - & 1.4 \\ 
N-hotspot-peak & 21:57:$09.932\pm0.003$,-69:40:$39.56\pm0.01$ & $25.6\pm0.8$ &
$0.7\times 0.5$ & $79\pm9$ & 30  & $+48$ & 0.7\\
N-hotspot-total & - & $50.5\pm 2$ &
- & - & - & - & 1.4 \\
S-hotspot-peak & 21:57:$04.112\pm0.003$,-69:41:$46.66\pm0.02$ & $68\pm3$ &
$0.5\times 0.3$ & $-3\pm22$ & 50 & $-80$ & 0.7 \\
S-hotspot-tip & 21:57:$03.926\pm0.007$,-69:41:$50.21\pm0.02$ & $5.9\pm0.4$ &
$0.8\times 0.3$ & $-73\pm8$ & 27 & $+7$ & - \\
S-hotspot-total & - & $223.9\pm 10$ &
- & - & - & - & 1.5 \\
\hline
\end{tabular}
\medskip
\begin{minipage}{\linewidth}
(1) Component name.  Three sub-components combine to make the flux
  density of each of the core and component D.  For each hotspot the
  total flux density within the 90$\mu$Jy beam$^{-1}$ contour and that
  in the brightest, most compact, component are reported separately.
(2) J2000 position. The
  relative coordinates have random errors as quoted, but
   the absolute accuracy is uncertain by 
   $\rm \pm (0^{s}\llap{.}007,0^{\prime\prime}\llap{.}04)$
   based on the phase referencing to B2353-686.
(3) Total 18-GHz flux density
  in component. (4)
Deconvolved FWHM of Gaussian major and minor axes. (5) Major axis position angle. 
(6) Polarization percentage. (7) Polarization position angle of the
electric vector.  See Figs.~\ref{fig:polplot}, \ref{fig:polplot-Nhot}
  and \ref{fig:polplot-Shot} to gauge the
 uncertainties in the polarization. (8) Spectral index between 17 and 19 GHz.  Uncertain due to
  inconsistencies with indices within the two sub-bands, presumably
  caused by the time-variable phase gradients. $\dag$ = significant improvement to residuals when
component included in the complex core structure,
but only marginal evidence that this flux density is associated with a separate component.
\end{minipage}
\end{table*}

\begin{figure*}
\centering
\includegraphics[width=6.2truein]{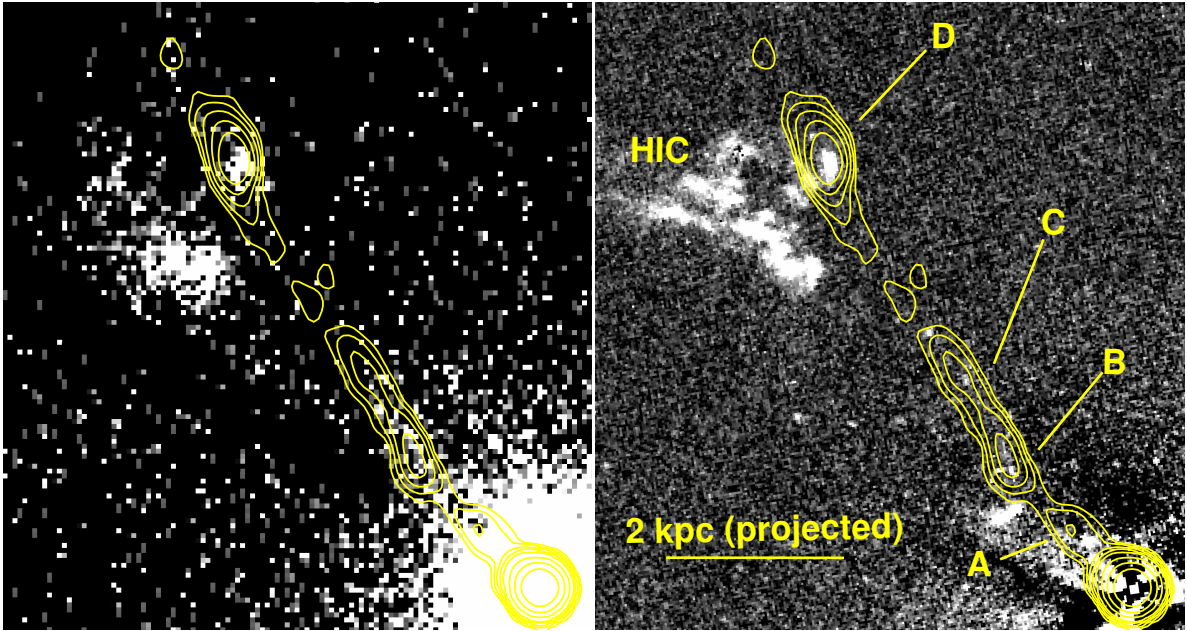}
\caption{ 
Left: 0.5-2.5~keV exposure-corrected unsmoothed image (0.0984-arcsec pixels)
with radio contours from Fig.~\ref{fig:mapjet}. The deficit of
counts on a line from the core to the S of the HIC is due to removal
of the readout streak from the core.  Right: \hst\ F606W
image (0.046-arcsec pixels) after galaxy
subtraction using the {\sc iraf ellipse} routine. The triangular
feature at the core is an artifact, but the inner V-shaped emission opening
out in the jet direction may be an edge-brightened ionization cone.
The optical emission just to the S of
knot B is believed to be real, and the patchiness near the core is
likely to be due to obscuration from dust.  Both the X-ray and optical
show emission from knot D, and
fainter emission tracking the jet mid-line at B and C.
}
\label{fig:xorad}
\end{figure*}

\subsection{Optical jet}
\label{sec:resultsopticaljet}

As seen in the right-hand panel of Figure~\ref{fig:xorad}, the
galaxy-subtracted \hst\ image shows not only emission in the region of
the HIC, interpreted as ionized gas by \citet{tadhunter87,
tadhunter88} and \citet{fosbury98}, but also emission associated with
radio knot D. The optical emission is more compact than radio knot D as a
whole, and the centroid gives an excellent match to the position of
sub-component D-peak.  We measure a flux density at $4.97 \times
10^{14}$~Hz of $2.5 \pm 0.2~\mu$Jy. The interpolated 18-GHz to optical
spectral index is then $\alpha_{\rm ro} \sim 0.76$.

Diffuse optical emission along radio components B and C is also seen,
totalling $2.0 \pm 0.2 \mu$Jy. Here the interpolated 18~GHz to optical
spectral index is $\alpha_{\rm ro} \sim 0.82$.  Although the galaxy
subtraction leads to some image artifacts, such as the triangular
feature at the nucleus, the patchiness of optical emission close to
the nucleus is most likely real and caused by dust obscuration.  There
is a significant feature immediately to the south of radio knot B.
The inner V-shaped emission opening
out in the jet direction may be an edge-brightened ionization cone.

\subsection{X-ray jet}
\label{sec:resultschandrajet}

The left-hand panel of Figure~\ref{fig:xorad} shows an X-ray excess
over background in the bright nucleus, knots B,C, and D, and the HIC.
The nuclear emission is the subject of a paper in preparation by
Momtahan et al.  In
this section we concentrate on the emission associated with the radio
knots, and results for the HIC are included in
Section~\ref{sec:resultsHIC}.

There are $\sim 100$ net counts ($0.3-7$~keV) from knot D over all
three exposures. The 18-GHz map is of a beam size well matched to
\chandra, and the centroid and concentration of X-ray counts make it
most plausible that the emission arises primarily from D-peak.  Using
a nearby region to measure the background (`nearbkg' in
Fig.~\ref{fig:regions}), and fitting using the {\sc xspec} CSTAT
statistic to a power-law model with Galactic absorption, we find a 1
keV flux density of $1.3 \pm 0.3$ nJy and a spectral index of
$\alpha_{\rm x} = 1.2 \pm 0.4$, which is in agreement with the more poorly
constrained result of \citet{young} for OBSID 1627 based on
hardness ratio.  The 0.3-5-keV luminosity is $1.3 \times 10^{40}$ ergs
s$^{-1}$.  OBSID 1627 used the S3 chip and occurred before the build
up of contaminants that have affected \chandra's low-energy response,
and so detected more counts than would be expected from a direct
scaling with exposure time.  Curiously there are twice the number of
net counts in OBSID 11530 (57) as in OBSID 12088 (28).  Although this
appears to be marginally significant it is not readily understood:
D-peak is resolved in the radio (Table~\ref{tab:radio}), and the X-ray
luminosity is sufficiently high that significant contamination from a
variable X-ray binary in the host galaxy is unlikely.  
The roll angles of the two observations differ by less than an
arcsecond, and knot D is sufficiently far from the readout streak (see
Fig.~\ref{fig:xorad}) that its counts are unaffected by the streak.
The discrepancy in counts is only significant above 1~keV, where it shows up
mostly where the spectral sensitivity is highest at 1--2~keV.
We disregard
any variability in what follows and attribute the average flux to
non-thermal emission from D-peak.  We note that the X-ray emission
from knot D is significantly harder than that in the HIC (see
Section~\ref{sec:resultsHIC}), and an attempt to fit a thermal model
finds an unrealistically high temperature of $kT \sim 2.3$~keV.

\begin{figure}
\centering
\includegraphics[width=3.25truein]{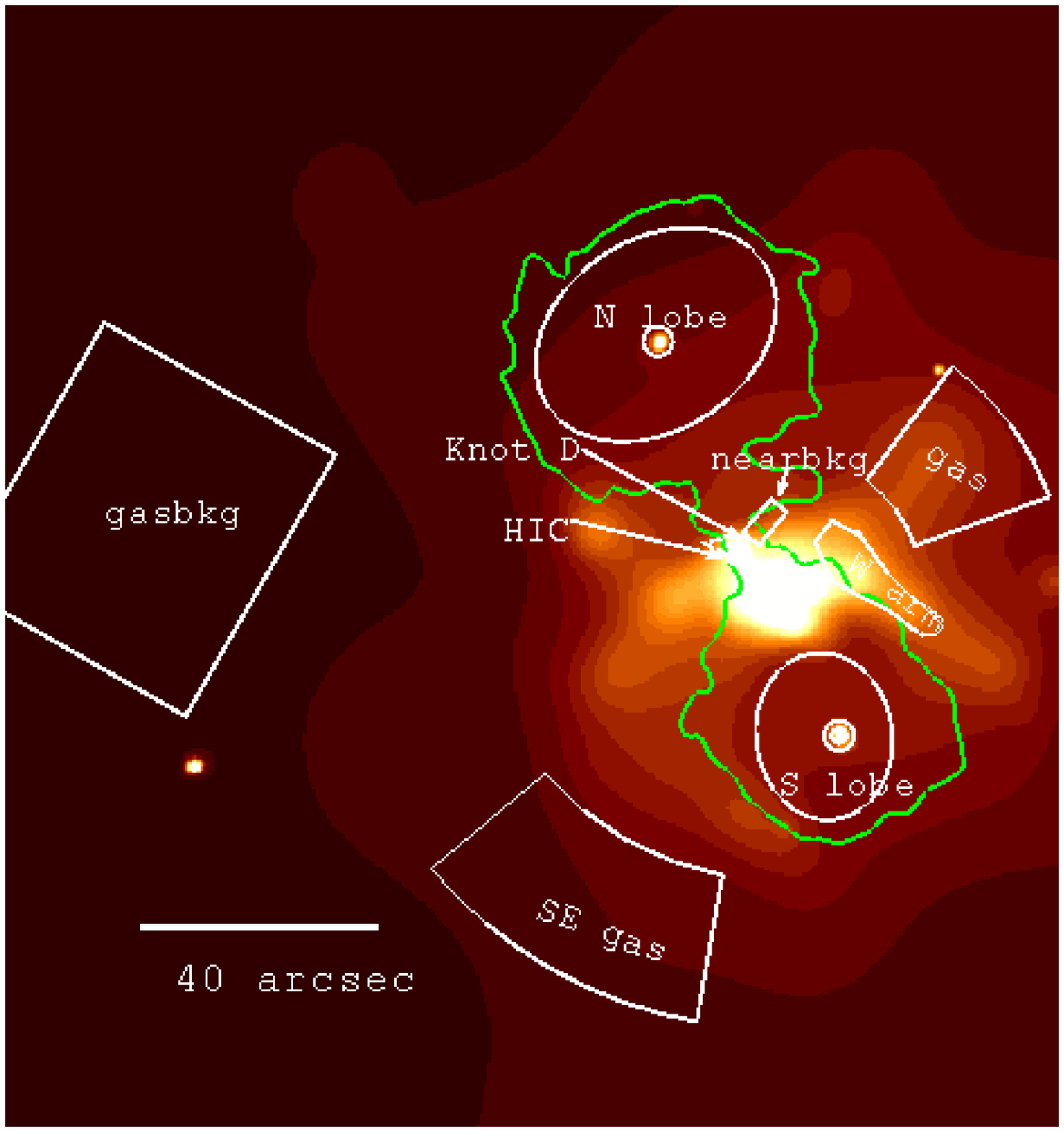}
\includegraphics[width=2.0truein]{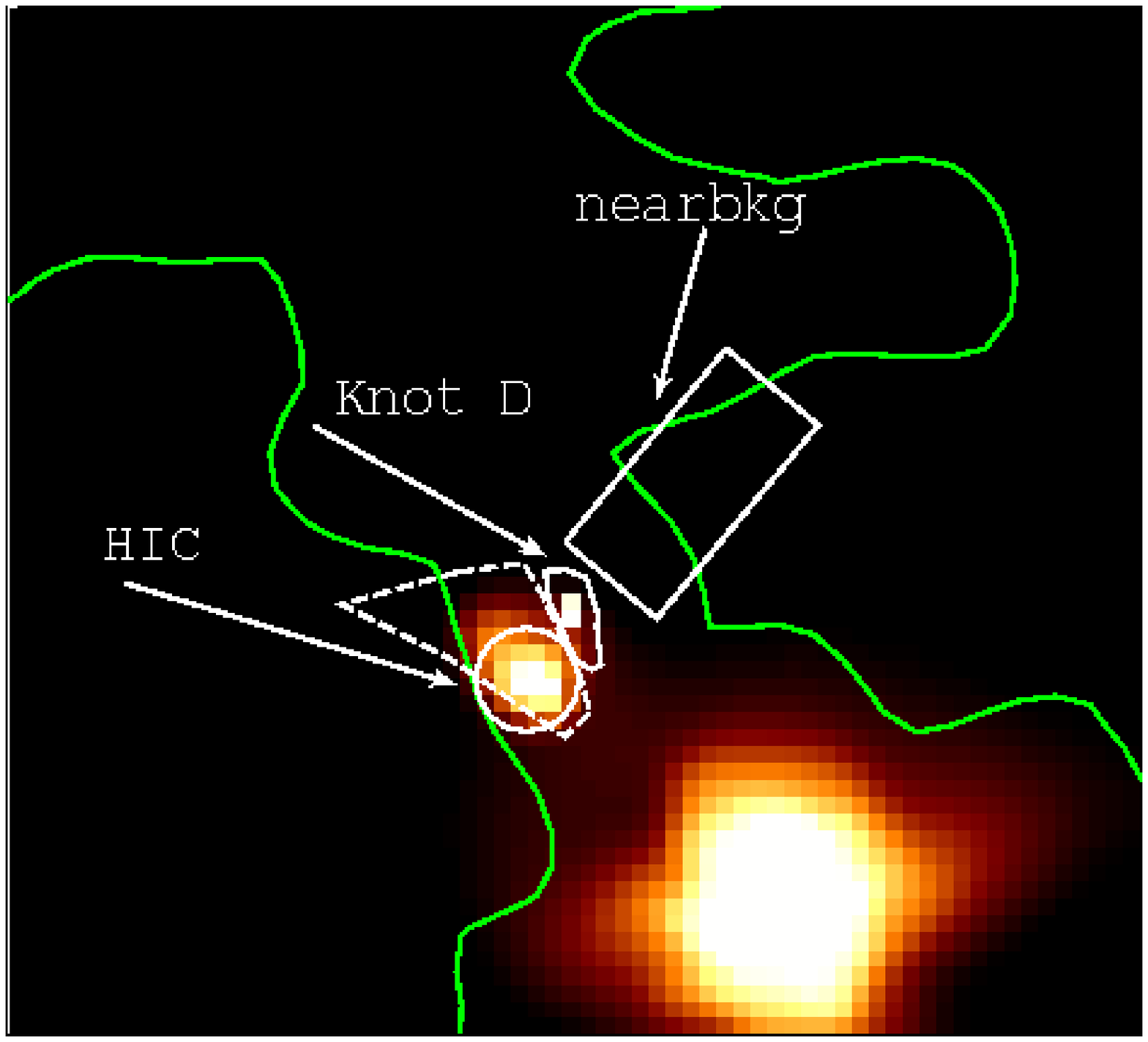}
\caption{ 0.5-2.5~keV exposure-corrected and adaptively smoothed
($3~\sigma$ threshold)
\chandra\ image of \source, outlining in white some of the regions
used for spectral extraction. The lowest radio contour from the map of Fig.~\ref{fig:xgas}
is shown in green.
 The
circles within the lobes mark the hotspot regions: in the case of the
S lobe the larger circle marks the excluded region for extracting lobe
counts, and the smaller circle is used to extract the peak of the
hotspot emission. Lower panel shows an expanded view of the inner regions: The HIC is marked by both a dashed
polygon and a circle covering its brightest emission to the SW.  }
\label{fig:regions}
\end{figure}

The net counts in knots B and C are measured using an adjacent
background region to the NW (so avoiding readout streaks to the SE in
OBSIDs 11530 and 12088).  We find only $45\pm10$ counts ($0.3-7$~keV)
over the three observations, which is too few for useful spectral
fitting.  An estimate of the luminosity of the emission
can be obtained through scaling values for knot D by the net counts
(see above), on the assumption that the spectral shape is similar.

\begin{figure}
\centering
\includegraphics[width=2.2truein]{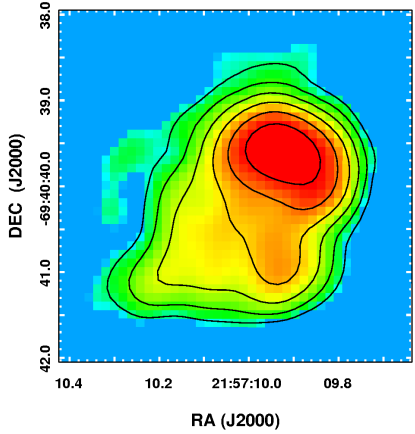}
\caption{
Expanded view of the northern hotspot from the map of
Fig.~\ref{fig:mapfull}.  Contours are at $(1,2,4,8,16,32)\times 0.11$
mJy beam$^{-1}$. The area within the $90\mu$Jy beam$^{-1}$ contour is 6.6
arcsec$^2$. }
\label{fig:mapNhs}
\end{figure}

\begin{figure}
\centering
\includegraphics[width=2.2truein]{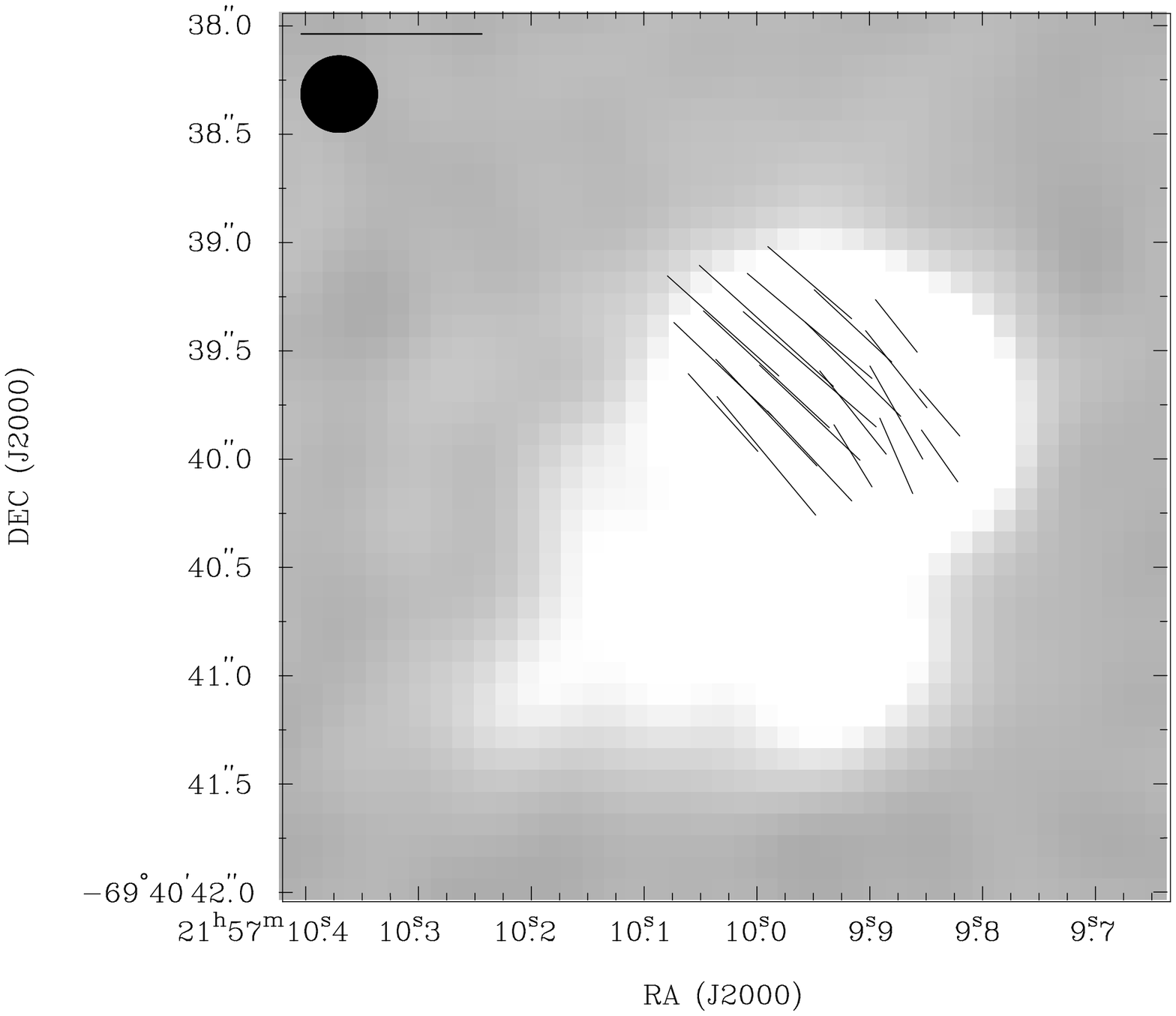}
\caption{
North hotspot shown with its fractional polarization E vectors.  The scale bar at top left
corresponds to 34.7 per cent polarization. The beam size is also shown
at the top left.
}
\label{fig:polplot-Nhot}
\end{figure}

\begin{figure}
\centering
\includegraphics[width=2.0truein]{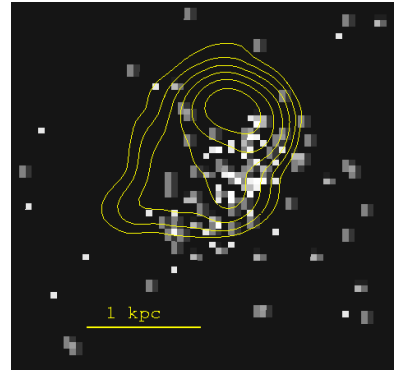}
\caption{
$0.3-5$ keV exposure-corrected unsmoothed image (0.0984-arcsec pixels,
log scaling, individual counts shown)
of the N hotspot, with radio contours from Fig.~\ref{fig:mapNhs}.
The X-rays are bright where the jet enters the hotspot, offset from the peak
of the radio emission.
}
\label{fig:NhotXR}
\end{figure}

\begin{figure}
\centering
\includegraphics[width=2.0truein]{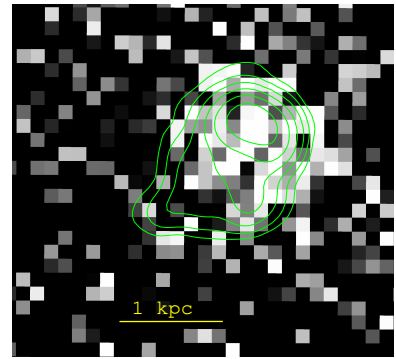}
\caption{
\wfi\ optical image (0.24-arcsec pixels,
log scaling) of the N hotspot, with radio contours from Fig.~\ref{fig:mapNhs}.
The optical is brightest close to the peak of the radio emission, but is
also somewhat elongated along the X-ray excess to the S.
}
\label{fig:NhotOR}
\end{figure}

\begin{figure}
\centering
\includegraphics[width=2.2truein]{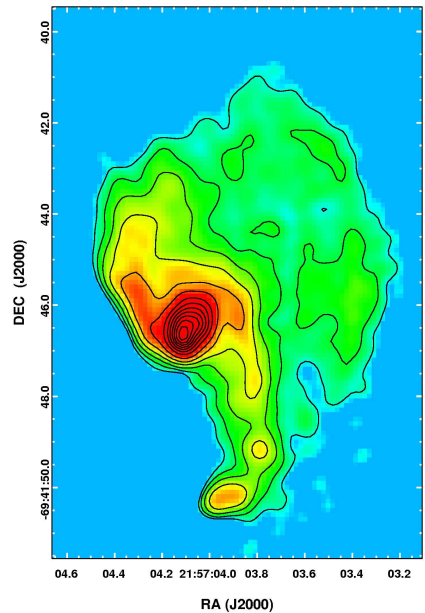}
\caption{
Expanded view of southern hotspot from the map of
Fig.~\ref{fig:mapfull}.  Contours are at $(1,2,4,8,16,32,64,96,128,160,192,224,256)
\times$ 0.095 mJy beam$^{-1}$. The area within the $90\mu$Jy beam$^{-1}$ contour is 38.3
arcsec$^2$.}
\label{fig:mapShs}
\end{figure}

\begin{figure}
\centering
\includegraphics[width=2.2truein]{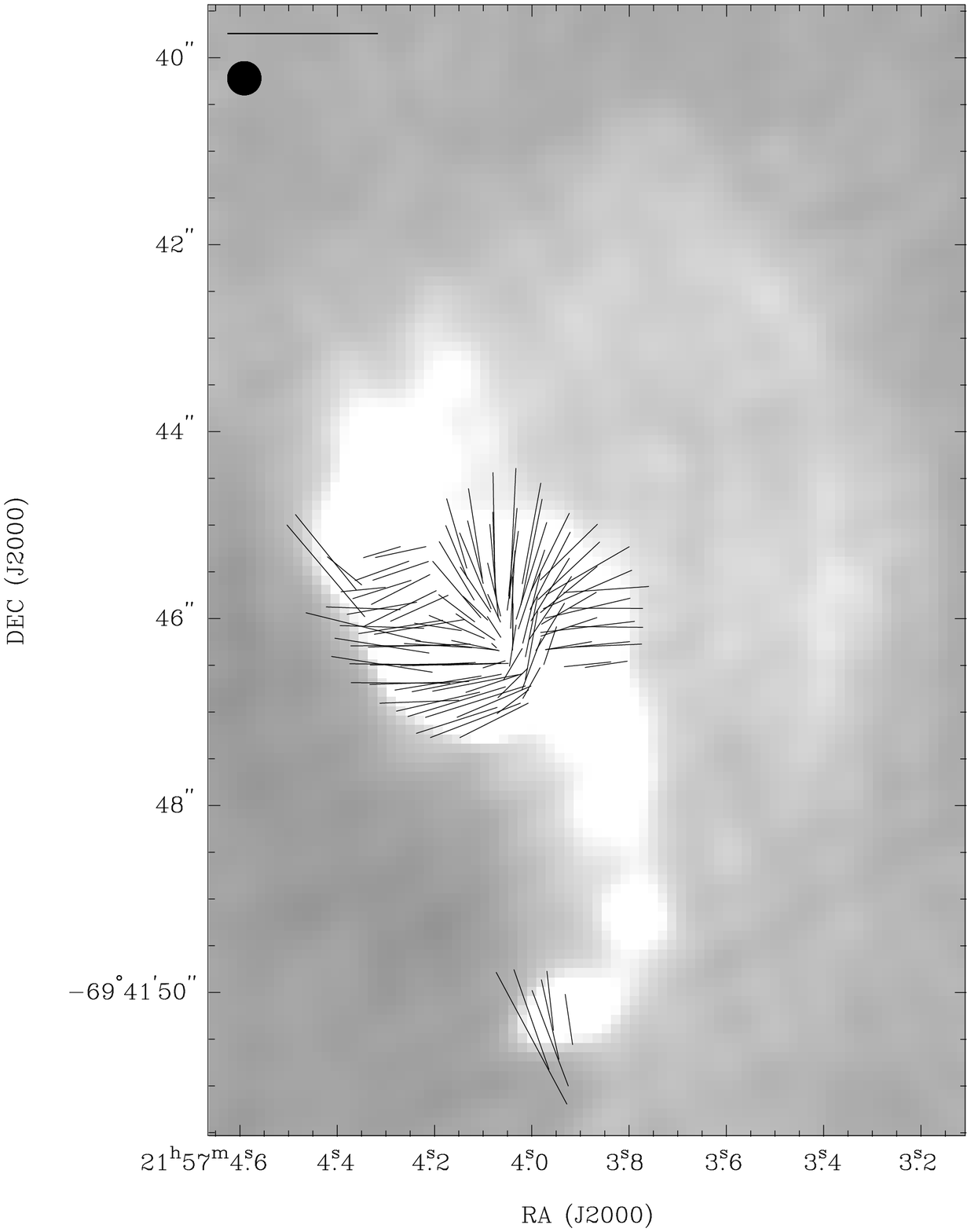}
\caption{
South hotspot shown with its fractional polarization E vectors.  The scale bar at top left
corresponds to 87 per cent polarization. The beam size is also shown
at the top left.
}
\label{fig:polplot-Shot}
\end{figure}

\begin{figure}
\centering
\includegraphics[width=2.2truein]{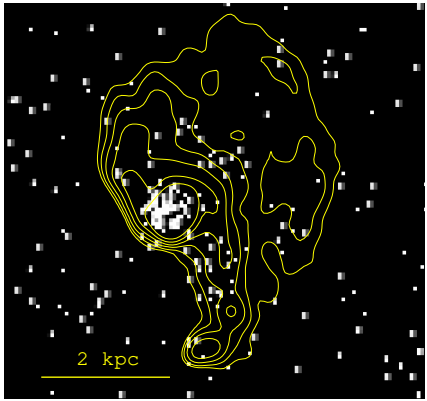}
\caption{
$0.3-5$ keV exposure corrected unsmoothed image (0.0984-arcsec pixels,
log scaling, individual counts shown)
of the S hotspot, with the six lowest radio contours from Fig.~\ref{fig:mapNhs}.
The X-rays are brightest at the peak of the
radio emission.
}
\label{fig:ShotXR}
\end{figure}

\begin{figure}
\centering
\includegraphics[width=2.0truein]{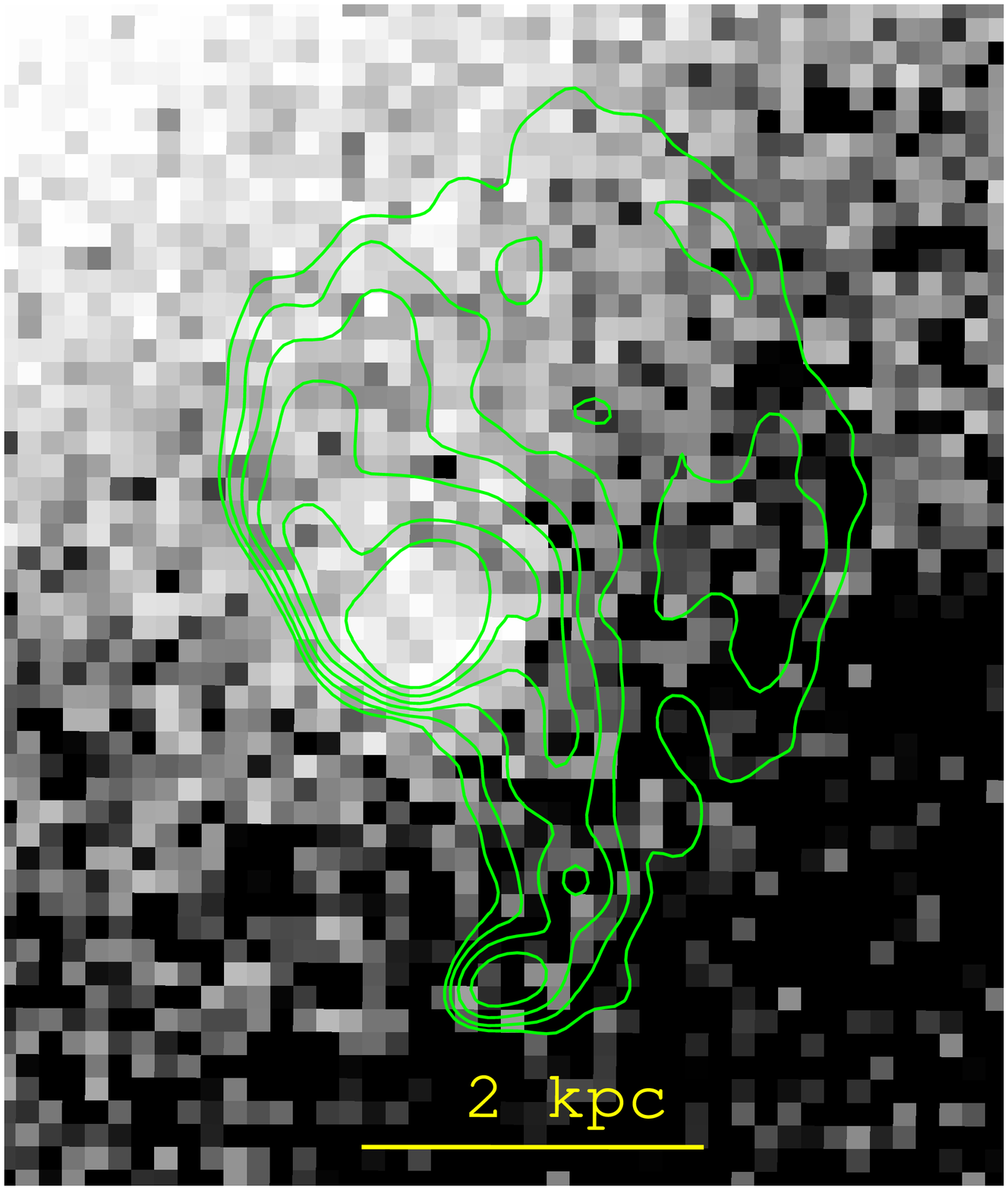}
\caption{
Optical image (0.24-arcsec pixels,
log scaling) of the S hotspot from the \wfi, with the six lowest radio
contours from Fig.~\ref{fig:mapNhs}.
Like the X-ray, the optical is brightest at the peak
of the radio emission.  The optical emission to the NE is galaxy light.
}
\label{fig:ShotOR}
\end{figure}

\subsection{Hotspots}
\label{sec:resultshotspots}

18-GHz intensity and polarization maps for the N and S hotspots are
shown in Figs.~\ref{fig:mapNhs}, \ref{fig:polplot-Nhot},
\ref{fig:mapShs} and \ref{fig:polplot-Shot}, and measurements appear
in Table~\ref{tab:radio}.
%The 18-GHz intensity and spectrum of the total N hotspot emission make
%a good extrapolation from the 4.7 GHz measurement of
%\citet{fosbury98}.  In the total S hotspot the intensity is a factor
%of $\sim 2$ higher than an extrapolation, but much of what we are
%labelling hotspot is seen as lobe emission in the lower-frequency
%data.  
While the total hotspot spectra are unusually steep at 18~GHz, as
commented also by \citet{fosbury98} for lower frequencies, the peaks
of the emission have more normal spectral indices of $\alpha_{\rm r}
\sim 0.7$ (Table~\ref{tab:radio}).
Our measurements for the peak of
the S hotspot are consistent with roughly one third of its flux
density being contributed by the VLBI components of \citet{young}
(unresolved in our beam), if we extrapolate between 1.4 and 18 GHz using
$\alpha_{\rm r} = 0.7$.

The X-ray emission in the N hotspot (Fig.~\ref{fig:NhotXR}) is clearly
offset from the radio peak: its centroid is 1.14 arcsec ($\sim
600$~pc) from the peak of the radio emission.
The X-rays are strong where the jet must be
entering from the S and where the radio emission is of steeper
spectrum than in the peak.  This is in marked contrast to the X-ray emission in the S
hotspot (Fig.~\ref{fig:ShotXR}) which shows good agreement between the
X-ray and radio peaks.  For the N hotspot we have fitted the $\sim
100$ net X-ray counts in a circle of radius 2.5 arcsec centered on the
radio peak to a power law using region `N lobe'
(Fig.~\ref{fig:regions}) as background.  The 1-keV flux density is
$1.0 \pm 0.2$~nJy and $\alpha_{\rm x} = 0.8 \pm 0.3$.  For the S
hotspot we have fitted the $\sim 100$ net X-ray counts in a circle of
radius 1.25 arcsec centered on the radio peak to a power law using
region `S lobe' (excluding a region of radius 2.5 arcsec around the
hotspot) as background (Fig.~\ref{fig:regions}).  The 1-keV flux
density is $1.25 \pm 0.25$~nJy and $\alpha_{\rm x} = 1.0 \pm 0.3$.

The \wfi\ observation finds optical emission associated with each of the N
and S hotspots (Figs.~\ref{fig:NhotOR} and \ref{fig:ShotOR}). Within
apertures of radius $\sim 1.5$~arcsec the flux densities at
$5.5~\times 10^{14}$~Hz are $2^{+2}_{-1}$ and $4^{+4}_{-2} \mu$Jy for
the N and S, respectively, with the errors dominated by the dispersion
when using different apertures and methods for background subtraction.
To determine if the hotspot optical emission is resolved we first
fitted Gaussians to nearby stars, finding an average FWHM of $2.43 \pm
0.02$ arcsec.  For the N hotspot, the fitted position is $\sim 0.4\pm
0.2$ arcsec to the south of the radio peak, and the deconvolved size
of the optical emission is $(2.4 \pm 0.4) \times (1.7 \pm 0.5)$
arcsec. The optical is thus mostly emission from the radio peak, with some
originating from the X-ray bright region to its south.  For the S
hotspot, the fitted position of the optical emission agrees within
errors with that of the peak of the radio emission, and the
deconvolved size is $(1.8 \pm 0.4) \times (1.1 \pm 0.6)$ arcsec.  Here
the X-ray and optical are both dominated by emission from the radio
peak.

\subsection{The HIC}
\label{sec:resultsHIC}

We have fitted models to the spectrum of the $\sim 429$ net X-ray
counts in an area of 5.3~kpc$^2$ at the HIC, using the on-source
(dashed) and background regions labelled `HIC' and `nearbkg',
respectively (Fig.~\ref{fig:regions}). Essentially all the detected
emission is below 3~keV.  A fit to a power-law model is only
marginally acceptable, and gives an unrealistically steep spectral
index of $\alpha_{\rm x} \sim 3.2$.  A thermal model fits well, with
$kT \sim 0.29$~keV.  What is striking, however, is the low metallicity
of the gas relative to solar ($Z/Z_\odot$), with values less than a
few per cent.  This is not true of other thermal components in the
field (see Section~\ref{sec:resultschandragas}).  Confidence contours
in $kT$ and abundance are shown in the left panel of
Fig.~\ref{fig:hictempabund}.  Abnormally low abundances can result
from neglecting to account for a component of unresolved low-mass
X-ray binaries (LMXB).  We have tested that in the HIC by applying the
method used by \citet{hump} and \citet{worralln4261} for other local
radio galaxies of adding in a thermal-bremsstrahlung model of free
normalization but fixed $kT = 7.3$~keV, the well-constrained best fit
for 15 nearby early-type galaxies from \citet*{irwin}.  The best-fit
0.3-5 keV luminosities are then $10^{41}$ and $7 \times 10^{39}$ ergs
s$^{-1}$ for the thermal gas and LMXBs, respectively.  Uncertainties
in the thermal bremsstrahlung normalization are large, and the
best-fit value is almost certainly an overestimate since the HIC
covers only a small area of the galaxy and the {\it total\/}
luminosities from LMXBs in 14 early-type galaxies are derived by
\citet{kimf} to lie in the range a few $\times 10^{39}$ to several
$\times 10^{40}$ ergs s$^{-1}$.  However, the effect of the inclusion
of a LMXB component on the fitted parameters for the thermal gas is
small (see right panel of Fig.~\ref{fig:hictempabund}), and the
normalization of the thermal component is reduced by less than 5 per
cent. In particular the gas metallicity is still constrained to be
low, and an X-ray binary component is not considered further.

\begin{figure}
\centering
\includegraphics[width=1.6truein]{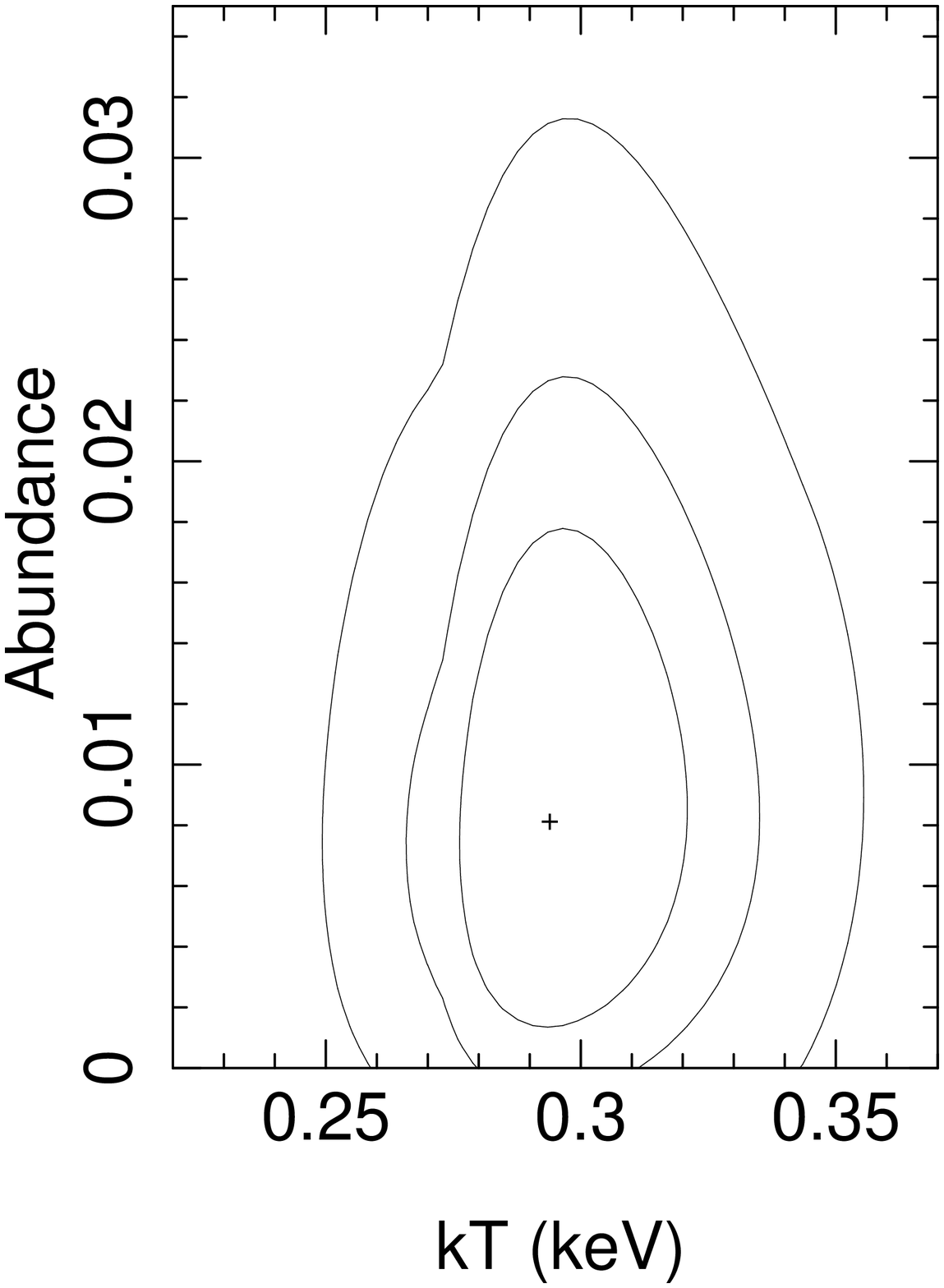}
\includegraphics[width=1.6truein]{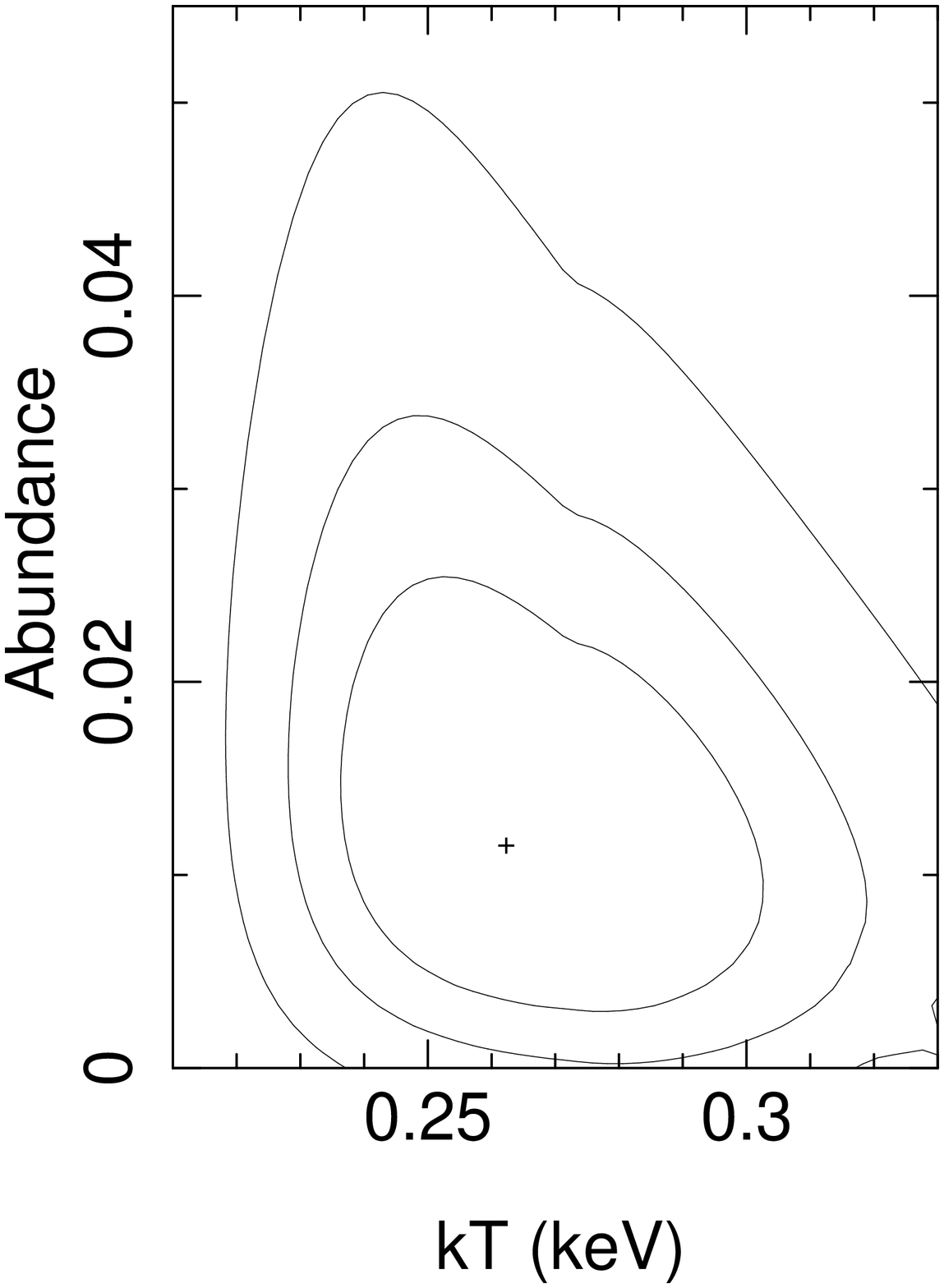}
\caption{
Confidence contours ($1\sigma$, 90\% and 99\% for two
  interesting parameters) for $kT$ and abundance ($Z/Z_\odot$)
for a thermal model fitted to the spectral data from the HIC. Left:
thermal model only. Right: a thermal bremsstrahlung model has been
added to allow for a contribution from X-ray binaries. }
\label{fig:hictempabund}
\end{figure}

A circle of radius 1.5 arcsec (0.85~kpc) centered on
RA=21:57:07.408. Dec=$-69$:41:16.53 contains 88 per cent of the HIC
net X-ray counts.  Parameter values for a fit to a thermal model are
essentially the same as in Fig.~\ref{fig:hictempabund} (left panel),
with the contours slightly larger due to fewer counts.  Since the
X-ray emission from the HIC extends further, particularly to the NE,
it is unlikely that the most intense emission arises from an
unresolved feature, and we treat 1.5~arcsec as the effective angular
radius of this complicated region in estimating gas density and
pressure in Section~\ref{sec:discussionHIC}. The bolometric luminosity
of the X-ray component in this circular extraction region is $3.3
\times 10^{41}$ ergs~s$^{-1}$.

\subsection{The large-scale gas distribution}
\label{sec:resultschandragas}

The most striking features of the larger-scale X-ray gas distribution
are arms around the lobes, particularly in the south, where there is a
clear gas cavity associated with the radio emission
(Fig.~\ref{fig:xgas}).  Here the lobe is embedded in gas, whereas the
N lobe appears not to be capped by bright gas at its N extremity.  The
overall appearance in the highest-surface brightness X-ray emission is
that of a truncated figure of eight.

\begin{figure}
\centering
\includegraphics[width=3.2truein]{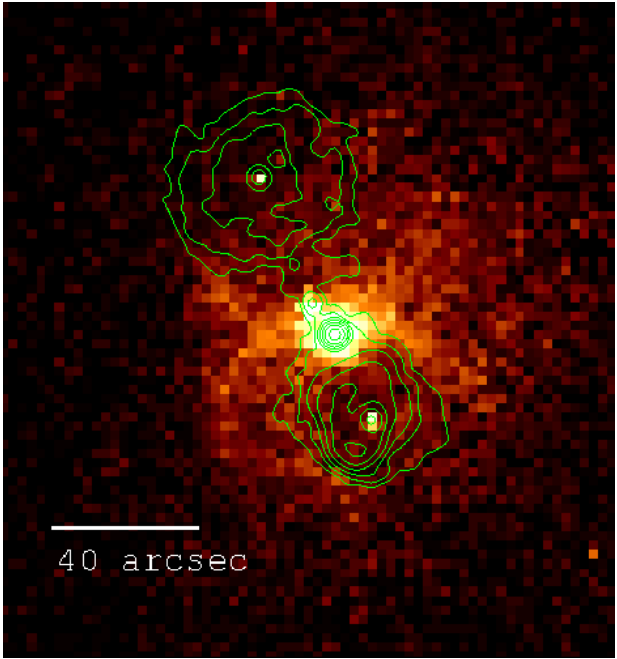}
\caption{
0.5-2.5~keV exposure-corrected unsmoothed \chandra\ image
in 2.5~arcsec bins. Contours are from 
the 4.74-GHz data of \citet{fosbury98}, mapped with
a 2.35-arcsec restoring beam.  Levels are
$(1,2,4,8,16,32,64,128) \times$ 3 mJy beam$^{-1}$.
 }
\label{fig:xgas}
\end{figure}

We have fitted spectral models to gas in regions indicated in
Figure~\ref{fig:regions}.  `W Arm' covers the brightest part of the
arm around the southern radio lobe.  The region labelled `gas' is a
section of a core-centered annulus which was chosen on the basis of
intensity and morphology to represent the location of undisturbed gas.
We have also applied spectral fitting to the weaker emission from a
core-centered annulus labelled `SE gas', lying to the south-east of
the southern lobe.  Checks were made to verify that results were
insensitive to the precise selection of a local background region
(taken to be no closer to the nucleus than `SE gas'), and results
presented here use the region labelled `gasbkg'. All the regions lie
fully on the S3 chip for OBSID 1627 and the I3 CCD for OBSIDS 11539
and 12088, and avoid the readout streak from the core and
serendipitous X-ray sources found using the {\sc ciao wavdetect} task.

The wings of the point-spread function (PSF) of the bright nucleus
supply position-dependent X-ray emission that it was found necessary
to model, since on-source and background regions were from different
off-axis angles.  As the central pixels of the nucleus suffer from
pile-up, we used measurements from the read-out streak (see Momtahan
et al., in preparation, for details) to characterize the spectrum that
was used in a 2-million-second simulation of the PSF with ChaRT at the
location of the nucleus in the long ACIS-I observations.  We then used
{\sc marx} to create an event file from which spectra could be
extracted and fitted using on-source and background regions at the
same locations relative to the optical axis as apply to the data from
\source. Fits of the spillover counts from the wings of the simulated
PSF to single-component power laws were acceptable for all the regions
described above.  The results were included in spectral models applied
to the \source\ data as fixed-parameter power-law components.
Including the PSF wing component caused the 0.3-5-keV luminosity of
the thermal gas emission to decrease by 24 per cent in `W Arm', but
the effect was smaller elsewhere.

\begin{table}
\caption{X-ray spectral fits to non-lobe gas emission}
\label{tab:xgas}
\begin{tabular}{lcccc}
\\
\hline
  \multicolumn{1}{c}{Region} &
  \multicolumn{3}{c}{Thermal} &
   $\chi^2$/dof\\
\cline{2-4} 
& $kT$ (keV) & $Z/Z_\odot$ &$N$ (Eq.~\ref{eq:apecnorm}, cm$^{-5}$)& \\
\hline
W Arm &
 $1.33^{+0.27}_{-0.11}$ & $0.31^{+0.60}_{-0.16}$ & $(2.5 \pm 1.0) \times 10^{-5}$
 & 19.4/17\\
gas & $0.95^{+0.05}_{-0.06} $ & $0.37^{+0.42}_{-0.17}$ & $(2.7^{+1.4}_{-1.2})
\times 10^{-5}$ &22.5/26\\
SE gas & $1.00 \pm 0.25 $ & $0.07^{+0.13}_{-0.06}$ & $(3.3^{+2.5}_{-1.2})
\times 10^{-5}$ &16.1/18\\
\hline
\end{tabular}
\medskip
\begin{minipage}{3.2truein}
Regions are from Fig~\ref{fig:regions} with `gasbkg' sampling the background.
Errors are $1\sigma$ for 2 interesting parameters.  Absorption
is consistent with the Galactic value and fixed. A fixed-parameter
power-law component based on simulations has been included to account
for the wings of the PSF from the nucleus.
\end{minipage}
\end{table}

All three regions described above then gave an acceptable fit to a
single-component thermal model (Table~\ref{tab:xgas}). When we take
into account the fact that the quoted errors are roughly equivalent to
90 per cent confidence on a single parameter, we see there is less
than 1 per cent chance of emission in `W Arm' being at the same
temperature as that in the `gas' region.  The `SE gas' also appears
cooler than `W Arm', although at lower confidence due to the reduced
intensity of this emission.

We checked how the results for the thermal gas would be affected by
including a contribution from unresolved LMXBs, as described in
Section~\ref{sec:resultsHIC}.  As compared with Table~\ref{tab:xgas}, for
`W Arm' $kT$ is only marginally affected and constrained to lie
between 1.19 and 1.48 keV, the abundances are now poorly constrained
at the high end, and the normalization (which is not used in
subsequent discussion) should be reduced by about 8 per cent (or more
if the gas abundance is truly anomalously high). In the other regions
the effect is also to decrease slightly $kT$ (to lie between 0.88 and
0.99 in the case of `gas') and the abundances are similarly affected.
Here the normalizations in Table~\ref{tab:xgas} are affected by a
negligible amount for the abundances quoted.  The conclusion that the
emission from `W Arm' is hotter than in the other regions
is therefore unaffected by a relatively small, but uncertain,
contribution from unresolved LMXBs.

\begin{figure}
\centering
\includegraphics[width=2.7truein]{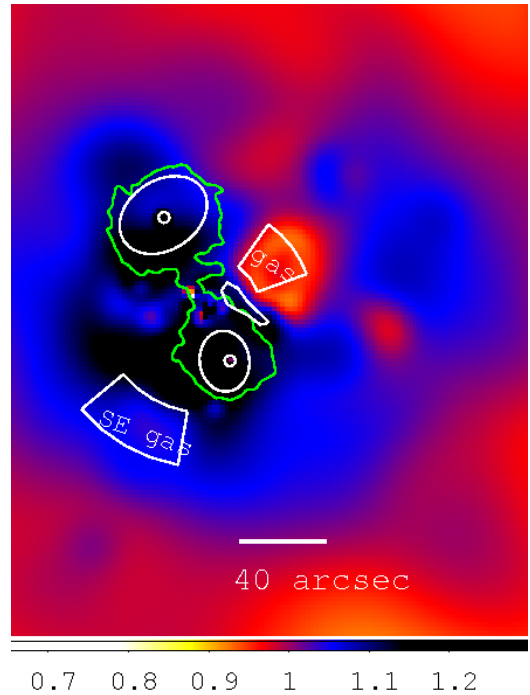}
\caption{ X-ray temperature map made following the prescription and
correlation between mean photon energy and temperature of
\citet{david}.  The colour bar is a rough indicator of $kT$ in
units of keV. Adaptive smoothing at 3$\sigma$ threshold has been
applied after blocking the data to a pixel size of 1.968 arcsec.
The lowest radio contour from the map of Fig.~\ref{fig:xgas}
is shown in green, together
with the locations of several regions from Fig.~\ref{fig:regions}.
 }
\label{fig:tmap}
\end{figure}

In an attempt to probe the overall temperature distribution of the gas
we have applied the method of \citet{david} to produce an image that
represents the gas temperature.  This method exploits the fact that
the L-shell line emission from thermal gas of $kT \sim 1$ keV is
blended into a broad peak whose centroid energy shifts to higher
values as the temperature increases, and is applicable for gas with
$kT$ roughly in the range 0.7--1.2~keV.  The method involves
producing an image of the mean energy of photons in the event file as
a function of position, and the image is subsequently smoothed for
display (Fig.~\ref{fig:tmap}).  A limitation of the method is that
since it works on individual events, no background is removed, and so
the background needs to be uniform in intensity and spectrum across
the field to avoid biasing the results.  In the case of \source\ this
is not the case for the contribution from the wings of the PSF, and so
we use the results here only in a qualitative way, and have applied
the calibration of mean energy to $kT$ of \citet{david} rather than
attempting a calibration for this particular field.  Nevertheless, the
results are enlightening.  In particular Fig~\ref{fig:tmap} shows that
the region labelled `gas', that was picked to represent undisturbed
gas purely on the basis of X-ray intensity and its location with
respect to the radio lobes, appears strikingly cool compared with
surrounding emission.  Indeed the temperature structure to the west of
the radio source appears to be more disturbed than that to the east.

\subsection{X-ray emission from the lobes}
\label{sec:resultslobes}

\begin{table*}
\caption{X-ray spectral fits to lobe emission}
\label{tab:xlobe}
\begin{tabular}{lccccc}
\\
\hline
  \multicolumn{1}{c}{Region} &
  \multicolumn{2}{c}{Power law} &
  \multicolumn{2}{c}{Thermal} &
   $\chi^2$/dof\\
\cline{2-3} \cline{4-5}
& $\alpha_{\rm x}$ & 1~keV normalization (cm$^{-2}$ s$^{-1}$ keV$^{-1}$) & 
$kT$ (keV) & $N$ (Eq.~\ref{eq:apecnorm}, cm$^{-5}$) \\
\hline
S Lobe & $0.90 \pm 0.24$ & $(7.3 \pm 1.1) \times 10^{-6}$ & - & - &27/23\\
N Lobe & $0.88^{+0.46}_{-0.72}$ & $(6 \pm 4) \times 10^{-6}$ &
$1.2^{+0.4}_{-0.2}$ & $(1.3 \pm 1.0
) \times 10^{-5}$ & 27/31\\
\hline
\end{tabular}
\medskip
\begin{minipage}{6in}
Regions are from Fig~\ref{fig:regions} with hotspots excluded and
 `gasbkg' sampling the background.
Errors are $1\sigma$ for 2 interesting parameters.  Absorption
is consistent with the Galactic value and fixed. Thermal fits are
insensitive to abundances which are fixed at $0.3 Z_\odot$.
\end{minipage}
\end{table*}

We have extracted the X-ray emission from the N and S lobes, using the
regions shown in Figure~\ref{fig:regions}, with the larger of the
inner circles used as hotspot exclusion zones.  Local background was
measured from the region marked `gasbkg', and the contribution from
the wings of the nuclear PSF was included in model fitting as
described in Section~\ref{sec:resultschandragas}.  When the emission
from the S lobe was fitted to a thermal model, the temperature was
high compared with other regions (Table~\ref{tab:xgas}), at $kT \sim
5$~keV.  A fit to a power law (Table~\ref{tab:xlobe}) is equally
acceptable, finding a spectral index in agreement with that of the
low-frequency radio emission of $\alpha_{\rm r}=0.8$
\citep{kuhr}. When thermal bremsstrahlung, to accommodate possible
emission from unresolved LMXBs, was included, its normalization tended
to zero. It is therefore most likely that the X-ray emission is
dominated by inverse-Compton scattering of the Cosmic Microwave
Background (CMB) radiation by the population of electrons responsible
for the radio emission.

The situation in the N lobe is more complex.  Again a fit to a
thermal model gave a higher temperature than other gas in the field,
but a power-law model gives a spectrum that is significantly steeper
than the radio.  The radio surface brightness is lower than in the S
lobe, and the extraction region needed to be larger to get $>500$ net
counts.  The larger region is expected to include significantly more
thermal gas surrounding the lobe than in the brightest part of the S
lobe where the thermal contribution appears to be negligible.  A
combination of a power law and thermal model not only gave an
improved fit ($\Delta\chi^2 = 5$) but more sensible best-fit
parameters given the overall source characteristics
(Table~\ref{tab:xlobe}).  

For application of an inverse Compton model
we have assumed that the X-ray spectral index matches the radio, in
which case the measurements of 1~keV flux density, with $1\sigma$
errors, are $4.6\pm0.3$ and $3.5\pm0.7$~nJy for the non-thermal
component in the S and N lobes, respectively.

\section{Discussion}
\label{sec:discussion}

\subsection{The jet morphology}
\label{sec:discussionjet}

The new 18-GHz data reveal the polarization structure and path taken
by the jet from the core on its way to the N hotspot.  As is often the
case in radio galaxies \citep[e.g.,][]{saikia, jorstad}, the core is
of low radio polarization, but after $\sim 0.35$ arcsec the
polarization has jumped to $\sim 20$ per cent, with an implied
magnetic field direction along the jet.  Through component~A the jet
continues in its VLBI direction, heading towards the HIC.  As pointed
out by \citet{tadhunter88} and \citet{fosbury98}, the direction
supports suggestions that the HIC is at least partially photoionized
by beamed emission from the inner jet.  We now know this inner jet
extends in the required direction for a projected distance of
$\sim 1.5$~kpc.  The inner
V-shaped emission seen in the \hst\ data (Fig.~\ref{fig:xorad}, right)
may be an edge-brightened ionization cone similar to that seen in, for
example, NGC\,1068 \citep{unger}.

The data then show the jet deflection to be complicated.  In component
B there is significant clockwise bending, accompanied by realignment
of the magnetic field to maintain its dominant alignment along the
jet.  The new angle would take the jet straight to the N hotspot,
except that the jet deflects anticlockwise in component C.  Diffuse
optical and X-ray emission are seen along knots B and C.  In component
D, the brightest jet knot, the jet deflects back clockwise to the
angle required to reach the N hotspot.  It passes through the
mis-oriented sub-component, D-peak, towards the leading edge of the
brightest emission in D.  D-peak has the curious property of being
misaligned relative to the rest of component D in position angle and
polarization angle (Table~\ref{tab:radio}).  Optical and X-ray
emission in knot D appear to be associated principally with D-peak,
giving an interpolated 18 GHz to optical spectral index of
$\alpha_{\rm ro} \sim 0.76$, which is flatter than the radio spectrum
integrated over all of knot D.  On leaving component D the jet is much
fainter and on a path towards the N hotspot. Further multiple
deflections are not required but cannot be excluded.

It is an interesting question as to why the jet takes a zig-zag path
through components A, B, C and D.  The discovery of optical emission
at the base of knot B, where the first such bending occurs, may be
relevant. The optical feature at the base of knot B has similar
surface brightness to the HIC, which lies adjacent to knot D where
more bending occurs.  If this comes from a finger of gas that is
pushing the jet and causing the deflection, spectroscopic measurements
of line widths should measure momentum transfer to the jet.
Jet deflection in knot C is less pronounced than in B and D, and there
is no bright feature in the HST image that might signal the presence
of excited gas there.

Our new radio measurements help to address the important, but
uncertain, angle to the line of sight, $\theta$, and bulk flow speed,
$\beta c$.  Use of the brightness ratio between approaching and
receding jets \citep[e.g.,][]{bridle}, $R_{\rm j}$, rests on the
assumption that internal and external factors are the same on each
side of the core.  This is not true for \source\ on large scales, since
the receding jet to the south has disrupted closer to the core and
produced a brighter hotspot and 18-GHz lobe.  If we nevertheless apply
the test to the inner jet at knots B and C, we measure $R_{\rm j} >
15$, from which we infer that $\beta \cos\theta > 0.45$ for an adopted
spectral index of $\alpha_{\rm r} = 0.8$.  The nuclear optical
spectrum, of intermediate ionization and exhibiting both forbidden and
permitted lines with broad wings, is described by \citet{tadhunter88}
as similar to that of Pictor A, a well-studied, somewhat more
powerful, radio galaxy at $z = 0.035$.  \citet{marshall} conclude that
Pictor~A's jet most likely lies at $10^\circ < \theta < 45^\circ$, as a
lobe-dominated object whose nuclear optical spectrum should be
accommodated by Unification models, and it is reasonable to adopt the
same argument for \source.  The radio core prominence of \source\ is
also comfortably within the range typical of broad-line objects, but
not extreme  (e.g., its value of $5
\times 10^{-3}$ can be compared with those for different source types
in figure 7 of \cite{hard}). 
When the constraint on $\theta$ is combined with the constraint
on $\beta \cos\theta$, the relativistic Doppler factor,
$\delta = \sqrt{1-\beta^2}/(1-\beta \cos\theta)$, is also constrained.  If
$\theta = 10^\circ$, $\delta$ for the jet is permitted to take on its largest
value, and largest range of values, being constrained to lie between
1.6 and 5.7 (corresponding to $\beta=0.46$ and 0.98,
respectively). 
At $\theta = 45^\circ$, the maximum allowed value of $\delta$ is
1.4, corresponding to $\beta=0.64$, and $\delta$ is lower for faster
bulk flows.  Under all these conditions, the counterjet is
constrained to have $\delta < 0.6$ ($< 0.5$ if $\theta = 45^\circ$).

\subsection{The jet knots}
\label{sec:discussionjetknots}

\begin{figure}
\centering
\includegraphics[width=2.0truein]{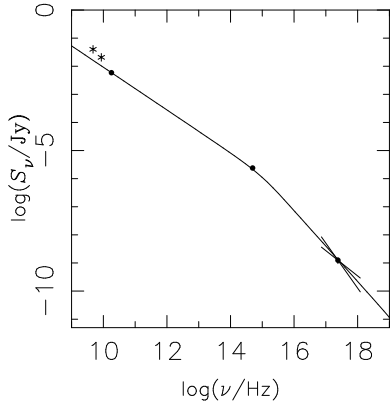}
\caption{
The radio--optical--X-ray spectral distribution in knot D-peak is
consistent with a broken power law arising from a single-component
electron spectrum emitting synchrotron radiation, with energy losses
being significant at the higher frequencies.
The two lowest-frequency radio points are from \citet{fosbury98} and
include emission over a larger region.
}
\label{fig:knotD}
\end{figure}

Figure~\ref{fig:knotD} shows the multiwavelength spectrum of knot
D-peak modelled with a broken power law.  The lower-frequency radio
measurements of \citet{fosbury98} are also plotted, but these contain
emission from the whole of knot D and some lobe emission.  The
spectrum of slope $\alpha_{\rm ro}=0.76$ must break if it is to
extrapolate as a single component to the X-ray: a similar result
applies to knots B/C.  This is as expected for synchrotron radiation
from a single power-law distribution of electrons if energy losses are
responsible for the break.  A simple energy-loss model predicts
$\Delta\alpha=0.5$, in which case the break arises at $\sim 8 \times
10^{14}$ Hz in knot D-peak, as plotted.  However, larger breaks are
often observed in low-power FRI jets \citep*[e.g.,][]{bm87, hard66b,
birkins}, in which case the break frequency will be somewhat higher.
The X-ray spectral index of $\alpha_{\rm x} = 1.2 \pm 0.4$ is
consistent with the extension of a broken power law, although
is insufficiently well measured to immediately rule out
an inverse Compton origin, for which the spectral slope
must match that of the radio synchrotron emission of the responsible
electrons.

We model the radio component D-peak as a cylinder of length 0.4 arcsec
and radius 0.05 arcsec (Table~\ref{tab:radio}). We find a
minimum-energy magnetic field of $B_{\rm me} \sim 46$~nT for a fully
leptonic plasma whose electron spectrum extends with spectral index of
$2\alpha_{\rm ro}+1=2.52$ down to a Lorentz factor of $\gamma_{\rm
min} =10$, and where the bulk relativistic Doppler factor is
$\delta$=1 \citep[for relevant equations see, e.g.,][]{wbrev}.  The
B/C region, modelled as a cylinder of length 3.3 arcsec and radius 0.1
arcsec, with a spectrum of $\alpha_{\rm ro}=0.82$ to connect the
18-GHz and \hst\ data points, gives $B_{\rm me} \sim 20$~nT.  If there
is significant relativistic beaming (see Section
\ref{sec:discussionjet}) and we treat this as a steady jet, the values
of $B_{\rm me}$ should be divided by $\delta^{(2+\alpha_{\rm
ro})/(3+\alpha_{\rm ro})}$.

The X-ray emission in quasar jets at small angle to the line of sight
is frequently modelled as arising from inverse Compton scattering of
CMB radiation that is boosted in the rest frame of a fast jet having a
minimum-energy magnetic field \citep[for a review see][]{wrev}.
For galaxy-scale jets, starlight can be more important than
the CMB for such scattering \citep*[e.g.][]{cel-pks0637, stawarz-starlight}. We have
fitted \source's starlight distribution, taken from our \wfi\ data, to a
S\'ersic profile, and derived the bolometric surface-brightness profile
using the total V-band magnitude given by \citet{devauc} and the
bolometric correction appropriate for a K2 star.  We deprojected
the profile to estimate the bolometric volume emissivity of starlight,
and integrated to find the local energy density as a function of
position in the galaxy.  We find that the X-ray
inverse-Compton contribution from starlight dominates over the CMB
at the deprojected position of knot D only if the viewing angle
$\theta > 7^\circ$.  We used
equation (7) of \citet{wrev}, modified to include both starlight and
the CMB, to calculate the beaming parameters required to
produce the X-ray emission.  We find that sufficient X-ray emission
is produced only if knot D lies at $\theta < 2^\circ$ (i.e.,
CMB-dominant scattering) with very large Doppler factors (e.g.,
$\delta \approx 44$ for $\theta =1.3^\circ$).  The values are
implausible for a radio galaxy and are well outside our estimates for
\source\ (Section~\ref{sec:discussionjet}).  While extreme bulk flow
and bending into the line of sight within a single knot might be
possible, X-rays are seen in other radio-bright jet regions in \source,
and a common origin for the X-ray would seem more plausible.

If we now proceed under the assumption that $\delta$ is close to
unity, inverse-Compton X-ray emission in the jet is dominated by
synchrotron self-Compton (SSC) emission.  At minimum energy, our modelling
of both D-peak and the B/C region produces too little X-ray emission
by a factor of $\sim 3000$.

We thus conclude that the jet X-rays arise from
synchrotron radiation.  The electrons emitting
at the spectral break, $\nu_{\rm br}$, in Figure~\ref{fig:knotD} are then of energy
$E \approx \sqrt{\nu_{\rm br}/(0.1 B_{\rm me})} \approx 4 \times
10^{11}$~eV, and lose their energy in $\tau \approx 0.13/(E B^2_{\rm
me}) \approx 150$ years.
Such a light travel time corresponds to an angular size of $\sim 0.1$
arcsec, which is comparable to the radio dimensions of D-peak, supporting
the idea of continuous acceleration occurring in a small region within
D-peak, presumably as the result of a strong shock.  A similar
argument applied to B/C finds that the electrons radiating at the
spectral break would lose their energy after travelling an angular
distance of about 0.35 arcsec. The observation of diffuse optical
emission over jet paths longer than this, particularly in knot B,
requires particle acceleration from diffuse regions or a series of
knots.

In contrast to FRIs, where \chandra\ synchrotron jets are common and
illustrate the need for particle acceleration remote from the cores
\citep*[][for a review]{worrall01, hard66b, wrev}, detections of X-ray
synchrotron emission in FRII radio-galaxy jets are rare. Thus the
requirement for particle acceleration within their flows (before the
terminal hotspots) is not well established, except in a few cases
usually associated with isolated bright knots
\citep*{wilson-pica, worr-3c346, kraft-3c403, hardcrost, kataoka}.  While
\source\ is not a tremendously powerful FRII source, it certainly does
not have the morphology of an FRI on either side of the core, making
the evidence for diffuse particle acceleration here of particular
interest.

\subsection{The hotspots}
\label{sec:discussionhotspots}

Our 18-GHz data provide good imaging and polarimetry of both hotspots.
While the overall spectral indices are steep, the peaks show flatter
spectra (Table~\ref{tab:radio}), in the range expected for particle
acceleration \citep[e.g.,][]{achterberg}.  The N hotspot is the
fainter.  In the radio its intensity and polarization structure are
reasonably simple.  It is interesting that the brighter hotspot is
that south of the core, where no jet emission is seen, and so
interpreted as the side oriented most away from the observer.  The
southern radio lobe is also brighter and smaller, presumably due to
enhanced interaction with the hot gas, as discussed in
Section~\ref{sec:discussiongas}, and we detect some lobe emission at
18 GHz despite the small beam and high frequency.  The radio structure
of the S hotspot is fascinating.  The appearance is that of a spiral
of plasma with longitudinal magnetic field.  VLBI components are
embedded in this spiral \citep{young}. The pattern extends to the
southern tip of the hotspot, where again polarization is strongly
detected.

\begin{figure}
\centering
\includegraphics[width=2.0truein]{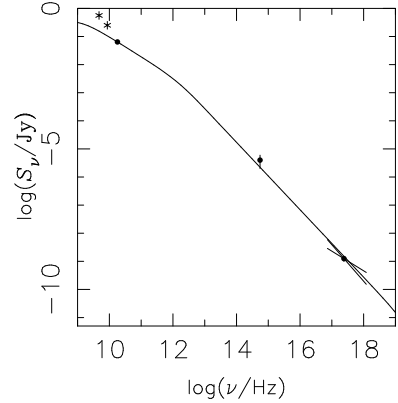}
\caption{
The 18~GHz, optical and X-ray measurements of the peak of the S hotspot
shown with a power law of
$\alpha_{\rm r} = 0.7$ breaking by $\Delta\alpha=0.5$ at
$\sim 10^{12}$~Hz.
The two lowest-frequency radio points are from \citet{fosbury98} and
include emission over a larger region.
}
\label{fig:shot}
\end{figure}

The peak of the S hotspot is bright in radio, optical and X-ray
(Figs.~\ref{fig:mapShs}--\ref{fig:ShotOR}).  If we interpret the
emission as synchrotron radiation in a broken power-law spectrum with
$\Delta\alpha = 0.5$ (Section~\ref{sec:discussionjetknots}), we find
that a power law of $\alpha_{\rm r} = 0.7$, breaking to $\alpha=1.2$
at $\sim 10^{12}$~Hz, is in reasonable agreement with the optical and
X-ray measurements, including the X-ray spectral index
(Fig.~\ref{fig:shot}).  We extrapolate the electron spectrum down to
$\gamma_{\rm min}=650$ \citep{godfrey} to find a minimum-energy
magnetic field of $B_{\rm me} \sim 25$~nT for $\delta = 1$, in which
case the predicted X-ray SSC emission falls below
observations by a factor of $\sim 100$.  
The shortfall in predicted X-ray emission is not
rectified by breaking the emission into the subcomponents of
\citet{young}, since the three subcomponents are all resolved with
VLBI and are distributed fairly evenly in flux density.  
In order to
explain the origin of the X-rays from the peak of the S hotspot as
SSC emission, the magnetic field strength would
need to be a factor of $\sim 13$ lower than the minimum-energy value to
provide sufficient particles for inverse Compton scattering without
exceeding the observed radio flux.  Such a
large departure from minimum energy makes a synchrotron origin seem
more likely; a better-constrained X-ray spectrum could provide a
definitive test.  While optical hotspot detections are still fairly
rare, in X-rays a synchrotron origin has been claimed for a number of
hotspots, increasing in fraction as source power decreases
\citep{hard}.  As a relatively low-power FRII source, \source\ is
consistent with such a trend.

If emission from the peak of the S hotspot is indeed synchrotron in
origin, the values above with equations from
Section~\ref{sec:discussionjetknots} give a lifetime of electrons
emitting at the spectral break, $\tau$, of $\sim 10^4$~yrs.  This is
about 10 times the light travel time across the hotspot, and
acceptable as magnetic fields help to confine particles within
hotspots.  \citet{perlman} use a synchrotron model for one of the
hotspots of 3C\,445 to argue that $\gamma_{\min}$ must be close to a
value of 1840, the ratio of proton and electron rest mass.  They find
that the optical emission is sufficiently intense to fit an unbroken
extension of the radio spectrum, requiring the magnetic field to be
relatively low to push an energy-loss break to high frequency.  This
is something that can be achieved for minimum energy by making
$\gamma_{\rm min}$ relatively high.  $\gamma_{\rm min}$ is the
uncertain factor having greatest influence on the value of $B_{\rm
me}$, and $B_{\rm me}$ is roughly proportional to $\gamma_{\rm
min}^{-(2\alpha - 1)/(\alpha+3)}$ for $\alpha > 0.5$ \citep[e.g.,
equation 57 of][]{wbrev}. We see from equations in
Section~\ref{sec:discussionjetknots} that for a fixed frequency of
spectral break, $\tau$ scales roughly as
$\gamma_{\min}^{(6\alpha-3)/2(\alpha+3)}$, i.e.,
$\gamma_{\min}^{0.162}$ for $\alpha = 0.7$.  That means that for
\source\ $\gamma_{\rm min}$ could be as low as $\sim 1$ without the
lifetime of electrons at the spectral break becoming shorter than the
light-crossing time of their emission region.

\begin{figure}
\centering
\includegraphics[width=2.0truein]{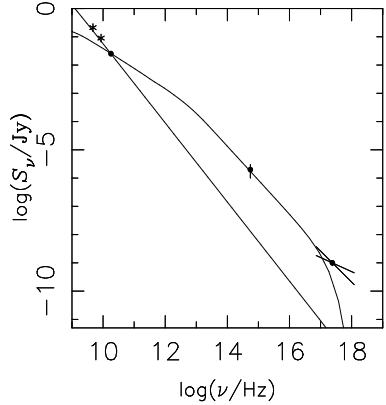}
\caption{
The 18~GHz, optical and X-ray measurements of the N hotspot.  The
flatter curve models the radio peak and optical emission as
synchrotron emission that must cut off before the spatially offset
X-ray emission.  The steeper curve is synchrotron emission in the
diffuse part of the hotspot which does not extrapolate to the X-ray
emission.  The two lowest-frequency radio points are from
\citet{fosbury98}.
}
\label{fig:nhot}
\end{figure}

While the most natural explanation for the X-ray emission from the S
hotspot is synchrotron radiation, an interpretation of the N hotspot
is more complex, as the X-ray emission is clearly offset from the
radio peak (Fig~\ref{fig:NhotXR}). In the radio peak a
synchrotron-emitting electron spectrum that also produces most of the
optical emission must cut off at energies less than $7 \times 10^{12}$
eV (in a minimum-energy magnetic field of $\sim 13$~nT) if it is not
to produce detectable X-ray emission at the radio peak
(Fig.~\ref{fig:nhot}).  We have deduced that the S hotspot accelerates
electrons to at least $10^{13}$ eV, and that the N hotspot is unable
to accelerate electrons beyond $7 \times 10^{12}$ eV.  Assuming a
common acceleration mechanism, it suggests the maximum energy to which
it is effective is of order $10^{13}$~eV rather than much higher, and we
are observing regional variations around this value.

As for the S hotspot, the overall N hotspot radio emission, including
on larger scales, is
too steep in spectrum for a continuation to higher energies to produce
the X-rays (Fig.~\ref{fig:nhot}).  For the X-ray to be synchrotron,
its spectrum must extrapolate back to optical and radio frequencies
with an index that is flatter than that in the most compact region of
the hotspot.  This would imply that the electron spectrum is less
affected by energy losses than in the more compact component:
that is difficult to understand unless particle acceleration
is of a different nature and well distributed within the
X-ray-emitting region.  The hotspot on the jet-side of Pictor~A has
been modelled as flat-spectrum X-ray-synchrotron emission, but there the
lack of a spatial offset with the peak of the radio emission
allows the X-rays to arise from compact VLBI
features, where it is argued that the electron spectrum to high
energies is flat because the particles esape from the small
acceleration zone before significant energy loss \citep{tingay-pica}.
Compatct VLBI features (yet to be detected) might be anticipated within
the radio peak of \source\, but not in the X-ray-emitting region.
We therefore conclude that if the X-rays from the N hotspot of \source\ are
synchrotron radiation, there are implications for the nature and
location of particle acceleration that have not previously been
modelled in a hotspot.

The X-rays in the N hotspot appear to be associated with the jet as it
approaches the peak location of particle acceleration, rather than the
hotspot itself.  Nothing similar is seen in the south, and it is
attractive to consider that the reason is associated with the fact
that the N hotspot is on the jet, rather than counterjet, side of the
source. The overall hotspot emission cannot be assumed to be strongly
affected by relativistic beaming, since the S hotspot, that is
associated with the receding jet, is the brighter, even taking only
the regions of concentrated emission, and hotspots are generally
considered to be relatively static, marking the location where the jet
terminates, and advancing only sonically into the external medium.
However, the jet plasma could have high relativistic bulk motion as it
enters the hotspot, and \chandra's detection of quasar jets, often
along their entire length, supports the idea that the flows of FRII
jets are not subject to the deceleration common in FRI sources
\citep[for a review see][]{wrev}.  The X-ray to radio ratio in the
X-ray-bright part of the N hotspot is similar to that in knot D, and
by analogy an explanation in terms of inverse Compton scattering of
CMB radiation would at first seem unlikely (see Section
\ref{sec:discussionjetknots}).  However, there is an essential
difference.  The radio spectrum in the hotspot (except in the peak,
offset from the X-ray) is steep spectrum.  This is observed within our
18~GHz data, and does not rely on lower-frequency measurements of
larger beam size.  The steep spectrum provides an excess of low-energy
particles which are those most effective in scattering the CMB to
X-ray energies.

We have modelled the X-ray-bright part of the N hotspot as a cylinder
of length 1.7~arcsec and radius 0.78~arcsec (see
Fig~\ref{fig:NhotXR}), with an electron spectrum of $\gamma_{\rm min}
=10$ and the correct slope to produce radio synchrotron radiation of
$\alpha_{\rm r}=1.4$ (Table~\ref{tab:radio}).  The minimum-energy
magnetic field for a Doppler factor of $\delta$=1 is found to be
$B_{\rm me} \sim 50$~nT, very similar to that in knot D (Section
\ref{sec:discussionjetknots}).  Scattering the CMB alone, we can
produce the observed X-ray emission if, for example, $\theta=8^\circ$,
and the bulk Lorentz factor and Doppler factor are both $\sim 7$.
Smaller angles to the line of sight are also allowed.  A value of
$\theta=8^\circ$ is somewhat uncomfortably small for a radio galaxy
(Section \ref{sec:discussionjet}).  However, as pointed out by
\citet{georg}, the bright hotspot synchrotron radiation itself
provides a photon field for the aproaching jet to scatter and beam in
its direction of travel before subsequent abrupt jet deceleration.  To
get some idea how large an effect this might be, we have estimated the
energy density of radio emission from the peak of the N hotspot at the
location of the X-ray emission, and used this as an additional
scattering field.  The availability of the additional photons relaxes
the constraint on $\theta$ only slightly, but we find $\theta\sim
10^\circ$ ($\delta \sim 5.7$) is now allowed.

An attractive feature of the jet being relatively highly beamed as it
enters the hotspot is that $B_{\rm me}$ reduces by a factor of
$\delta^{0.77} \sim 4$ from the unbeamed value of $\sim 50$~nT, and
now matches our minimum-energy estimate for the peak of the N hotspot
(13 nT) where the field is highly ordered
(Fig.~\ref{fig:polplot-Nhot}).  A slight concern is that the X-ray
spectral index would be expected to to be steep and match that in the
radio.  Our X-ray measurement, although poorly constrained, is
somewhat flatter (Sec.~\ref{sec:resultshotspots}).  What is clear is
that the N hotspot region, with its offset radio and X-ray emission,
has the observational advantage that the X-ray emission is not bright
at the hotspot itself, where it can confuse the study of the jet
emission.  Further sensitive radio and X-ray measurements will allow
more detailed modelling and should advance an understanding of jet
termination, since the N hotspot region is well-suited to sensitive
radio and X-ray observation in terms of both angular size and dynamic
range.

\subsection{The HIC}
\label{sec:discussionHIC}

We calculate the density, pressure, and cooling time of the
X-ray-emitting gas in the HIC following equations in \citet{wbrev}.
The emission measure is defined in terms of the normalization factor
returned by {\sc xspec}, $N$, as

\begin{equation}
10^{14} N = {(1+z)^2 \int n_{\rm e} n_{\rm p} dV \over 4\pi D_L^2},
\label{eq:apecnorm}
\end{equation}

\noindent
where $n_{\rm p}$ is the density of hydrogen nuclei, $n_{\rm e}$ is
the electron density, $V$ is volume, and $D_L$ is luminosity distance.
$n_{\rm e} \sim 1.18 n_{\rm p}$ due to the presence of He nuclei (the
factor is very weakly dependent on the abundance of heavier elements).
This means that for a spherical region of uniform density and angular
radius $\theta$, the density and pressure\footnote{1 Pa = 10 dynes cm$^{-2}$} are given by

\begin{equation}
(n_{\rm p}/{\rm m}^{-3}) = 8.5 \times 10^{8} (1+z)^2 \sqrt{(N/{\rm cm}^{-5}) \over
(D_L/{\rm Mpc}) (\theta/{\rm arcsec})^3},
\label{eq:dens}
\end{equation}

\begin{equation}
(P/{\rm Pa})
= 3.6 \times 10^{-16} (n_{\rm p}/{\rm m}^{-3}) (kT/{\rm keV}).
\label{eq:pressure}
\end{equation}

\noindent
For a monatomic gas the cooling time is then given by

\begin{equation}
(\tau/{\rm Myr}) = {2.9 \times 10^{48}  (kT/{\rm keV}) (N/{\rm
cm}^{-5}) (D_L/{\rm Mpc})^2 \over (1+z)^2 (n_{\rm p}/{\rm m}^{-3}) (L/{\rm ergs~s}^{-1})},
\label{eq:cooltime}
\end{equation}

\noindent
where $L$ is the {\it bolometric\/} luminosity of the thermal component.  For
convenience we express $N$ and $L$ in cgs units since this is how they
are returned by {\sc xspec}.  The sound speed is

\begin{equation}
(c_{\rm s}/{\rm kpc~Myr}^{-1}) = 0.54 (kT/{\rm keV})^{0.5}.
\label{eq:soundspeed}
\end{equation}

If we assume that the brightest X-ray emission from the HIC fills a
sphere of radius 1.5~arcsec (diameter 1.7~kpc) with uniform density, results of the
spectral fit together with Equations~\ref{eq:dens} and
\ref{eq:pressure} give a density of $n_{\rm p} =(7.8\pm0.7)\times 10^5$~m$^{-3}$
and pressure of $P = (8\pm2) \times 10^{-11}$~Pa.  The
cooling time (Eq.~\ref{eq:cooltime}) is then $15\pm4$~Myr, and
represents a maximum lifetime of the gas, unless there is a
significant heat source, for which the jet would be a strong
contender.  Since \source's age may be of order 15~Myr, we
cannot differentiate between a situation in which the X-ray emission
of the HIC became bright at the onset of the radio source, or more
recently.

However, the sound crossing time presents a more severe constraint.
In the absence of confinement by an outer atmosphere, the gas should
have suffered adiabatic expansion in roughly its sound crossing time
of $5.8\pm0.2$~Myr (Eq.~\ref{eq:soundspeed}).  While there is a hotter
X-ray emitting atmosphere surrounding the HIC gas
(Section~\ref{sec:resultschandragas}), the volume is large and the
surface-brightness contrast with the HIC gas is much too low for this
component to provide the necessary confinement of the HIC.  We thus conclude
that the HIC X-ray gas is relatively young.

The HIC is only one of the regions adjacent to locations of jet
deflection (see Sec.~\ref{sec:discussionjet}).  We are not able to
determine if X-ray-emitting gas binds the optical feature to the south
of knot B, due to the bright readout streak on the deepest X-ray
exposures (Fig.~\ref{fig:xorad}).  There is no evidence of an optical
feature where the jet deflects between knots B and C.  Momentum
transfer from a deflected jet is a natural reason to expect optical
filaments, but if the cool gas must be bound by a hot medium, short
lifetimes may be the reason that HIC-like features are not more
common.

It is interesting that short-lived blobs of X-ray-emitting gas are
also observed within a filament lying adjacent to radio emission
in the Northern Middle Lobe (NML) of Centaurus A \citep{kraft-nml},
where it is suggested that they result from cold gas that has been
shock heated by direct interaction with the jet.  There too the
elemental abundances are low, and it is suggested that this is an
artifact of unresolved temperature structure \citep[see also][]{kim}.
The same is possible for \source, which would benefit from deeper
X-ray observations.  It is also possible that the inferred abundances
are smaller than the true values if the plasma is photoionized: e.g.,
\citet{young01} find low abundances for
collisionally-ionized
models in NGC~1068, while \citet{kinkhabwala} find consistency with solar
abundances using photoionization models.

Further understanding of the HIC should result after
our approved spatially-resolved optical spectroscopy
has been carried out.  Some of the optical
emission may be the result of photoionization from the AGN
\citep{tadhunter87, alighieri}, but the
jet deflection in the vicinity of the HIC that is seen in this
first detailed mapping supports the idea that
direct interaction with the jet has been key to the formation of
the HIC \citep{tadhunter88, tingay, fosbury98}.  The linear optical structures of the HIC, as seen with \hst, give the
appearance of `scars', where the jet has passed through and `ignited'
particular gas clouds, where energy is still being dissipated in the
form of optical emission lines.

\subsection{Lobe inverse Compton emission}
\label{sec:discussionlobes}

If the regions from which lobe X-ray emission is extracted are
modelled as prolate ellipsoids, they have volumes of $1.4 \times 10^3$
and $4.25 \times 10^3$ kpc$^3$ for the S and N, respectively.  The
corresponding 4.74-GHz flux densities are 4.44 and 1.98 Jy.  The
minimum-energy magnetic fields for a fully leptonic plasma whose electron
spectrum extends with spectral index of $2\alpha_{\rm r}+1=2.6$ down
to a Lorentz factor of $\gamma_{\rm min} =100$ are $B_{\rm me} \sim
3.1$ and $1.9$~nT, with total pressures in the relativistic leptons and
field of $2.8 \times 10^{-12}$ and $1.0 \times 10^{-12}$ Pa,
respectively.  
The minimum-energy magnetic fields are slightly larger if
an energy density in relativistic protons equal to that in leptons is included
(3.8 and 2.3~nT, respectively), with corresponding pressures of
$4.0 \times 10^{-12}$ and $1.5 \times 10^{-12}$ Pa.

To produce the power-law X-ray emission found in our
spectral fits from inverse Compton scattering of the CMB, the field
values are instead $B = 1.0$ and 0.7~nT , with pressures in
relativistic leptons and magnetic field of $1.15
\times 10^{-11}$ and $3.2 \times 10^{-12}$ Pa for the S and N lobe,
respectively.  This level of decrease of the magnetic field strength
from minimum energy (assuming no relativistic protons) is commonly
estimated in radio galaxies for which
the presence of lobe inverse Compton emission is inferred \citep{croston}
, and leads to a significant increase in pressure over the
minimum-energy value. Since the particles dominate the magnetic field
in energy density, the pressure scales directly with the relativistic
particle content.
Thus if the energy density in relativistic protons equals that in leptons,
the corresponding pressures are higher at  $2.3
\times 10^{-11}$ and $6.4 \times 10^{-12}$ Pa.
The lower pressure in the N lobe is consistent
with a morphology that suggests less ability to drive a shock into the
surrounding gas, consistent with the absence of features
bright enough to produce a detectable shock spectral signature.

\subsection{The gas structure and radio cavities}
\label{sec:discussiongas}

The hot gas distribution in \source\ is seen sufficiently well in the
new \chandra\ observations to reveal some of its complexities.
The intensity map (Fig.~\ref{fig:xgas}) shows the radio lobes to have evacuated
cavities, and the arms of increased X-ray emissivity
are suggestive of shock heating by the lobes.
The temperature map (Fig.~\ref{fig:tmap}) suggests
large-scale shifting of gas. This is particularly noticeable to the
west of the radio source.  The low-frequency (843~GHz) radio map 
presented by \citet{tadhunter88}, that represents the location of
radio plasma where the particles have aged beyond those in maps presented
here, show the lobes to extend considerably to the west, but 
cut off sharply in the east.  Thus there is reason to believe that gas in our region
labelled `gas' in Figures~\ref{fig:regions} and \ref{fig:tmap} has been
displaced in the past by radio plasma.  As it is cooler than
its surroundings, it may now be relatively undisturbed.

The `gas' region is a 35-degree pie
slice of an annulus (fraction $f = 35/360$) of inner and outer radii
$\theta_1 = 24$ and $\theta_2 = 48$ arcsec, respectively,
centered on the nucleus of \source.  If the true emission volume is spherical of
radius $\theta_{\rm t}$ centred on the nucleus, the line of sight
volume, $V$, is given by

\begin{equation}
V = f D_{\rm A}^3 {4 \pi \over 3} \left[ \left(\theta_{\rm t}^2 -
\theta_1^2\right)^{3/2}
-\left(\theta_{\rm t}^2 -
\theta_2^2\right)^{3/2} \right].
\label{eq:vol}
\end{equation}

\noindent
Assuming uniform density between $\theta_1$ and $\theta_2$, the
modified version of Eq.~\ref{eq:dens} that gives the density in terms
of the observables is

\begin{eqnarray}
(n_{\rm p}/{\rm m}^{-3}) &=&  8.5 \times 10^8 (1+z)^2 
\sqrt{(N/{\rm cm}^{-5}) \over
f (D_L/{\rm Mpc})}
 \nonumber \\
&& \times
 \left[ \left(\theta_{\rm t}^2 -
\theta_1^2\right)^{3/2}
-\left(\theta_{\rm t}^2 -
\theta_2^2\right)^{3/2} \right]^{-1/2},
\label{eq:dens2}
\end{eqnarray}

\noindent
where all values of $\theta$ are in units of arcsec.  Pressure is
given by Eq.~\ref{eq:pressure}.  We adopt $\theta_2 = \theta_{\rm t}$,
since the emission is very much fainter beyond $\theta_2$, and in any
case the density and pressure cannot be higher than the values given
under this assumption.  We have also subdivided the annulus in two in
radius, and checked that spectral fits are consistent within errors,
suggesting no large gradient in temperature or abundance of the gas
within the region.  Using results in Table~\ref{tab:xgas}, the
pressure in this region is $1.7 \times 10^{-12}$~Pa.  A similar
argument applied to gas from the `SE gas' regions finds $1.2 \times
10^{-12}$~Pa.

The pressure in leptons and magnetic field required to produce the
lobe inverse Compton X-ray emission (Section 4.5) is higher than these
values for the external gas.  This is not unexpected since the the
lobes have created cavities in the gas.  The lobes are therefore
expected to drive shocks into the external gas, and be surrounded by
cocoons of heated dense gas which should be observable in X-ray
observations through increased intensity and temperature.  However,
the cocoon emission is diluted by foreground and background emission
along the line of sight, and in most cavity sources no evidence of
such shocked gas is seen \citep{mcnamara}.  \source, however, displays
prominent arms of increased emission around the lobes
(Fig.~\ref{fig:xgas}).  We have carried out spectral fitting for the
brightest section of the western arm, and the justification for
associating gas here (the `W Arm' region) with a section of such a
cocoon lies not only in its increased X-ray intensity but also the
evidence that the emission here is hotter (Table~\ref{tab:xgas}) than
in the regions sampled as being characteristic of unshocked gas.

The pressure ratio between the southern lobe (based on inverse Compton
emission for no relativistic protons) and the external gas is 6.  This
increases to 12 if the energy density in relativistic protons matches
that in leptons.  Assuming the shocked gas is in pressure balance with
the lobe, these ratios are those between shocked and unshocked gas.
Using the Rankine-Hugoniot conditions for a strong shock in a
monatomic gas \citep[e.g., eq.~49--51 of][]{wbrev}, this pressure
ratio relates to the Mach number of the shock via $(5 {\cal M}^2 -
1)/4$.  We thus infer a Mach number in the range 2.2 to 3, depending
on the relativistic proton content.  The temperature ratio of shocked
and unshocked gas for such a range of Mach number is 2.3 to 3.7, which
could explain the increased temperature seen in the `W Arm' region
once cooler gas along the line of sight to the cocoon is taken into
account.  The gas contains sufficient structure that it requires
complex spatial models to describe fully its distribution.  In a
forthcoming publication we will address this in more detail.

We have estimated the cavity and 1.4-GHz radio powers,
$P_{\rm cav}$ and $P_{1.4}$, respectively, as defined by
\citet{cavagnolo}.  $P_{1.4}$ is $7.7 \times 10^{41}$ ergs s$^{-1}$
\citep{kuhr}.  $P_{\rm cav}$ is raised by a factor of three through
our use of the dynamical time based on ${\cal M} = 3$ and the external
sound speed, rather than the buoyant rise time.  Despite this, our
estimate of the combined cavity power for the two lobes of $P_{\rm
cav} \sim 3 \times 10^{43}$ ergs s$^{-1}$ falls well below the
correlation with radio power of \citet{cavagnolo} ($\sim 2.6 \times
10^{45}$ ergs s$^{-1}$) and the earlier correlation of \citet{birzan}
($\sim 3.2 \times 10^{44}$ ergs s$^{-1}$).  However $P_{\rm cav}$
always underestimates the time-averaged jet power, because it ignores
work done driving gas and lobe plasma motions and heating shocked gas,
and these terms are likely to be particularly important given the
strong shock observed in \source.  We have estimated the additional
energy based on the southern lobe, and doubled the results.  We find a
kinetic energy of shocked gas, ${1\over 2} M_{\rm shock} ({2\over3}
v_{\rm shock})^2$ for ${\cal M}=3$, of $\sim 2.4 \times 10^{44}$ ergs
s$^{-1}$.  The mass in shocked gas, $M_{\rm shock}$, uses the region
`W Arm' as a representative cut through a spherical shell of inner and
outer radii 19 and 25 arcsec, respectively, whose density is found
using Eq.~\ref{eq:dens2} and the entry in Table~\ref{tab:xgas}, and we
adopt a shock speed, $v_{\rm shock}$, of three times the sound speed
in ambient gas represented by region `gas'.  While the surface
brightness peaks in `W Arm', the shocked region appears broader
elsewhere, and we assume the factors balance out in our estimate of $
M_{\rm shock}$.  A third contribution to the time-averaged jet power
($\sim 7 \times 10^{43}$ ergs s$^{-1}$) is from heating the shocked
gas from $kT = 0.95$ to 1.33 keV (Table~\ref{tab:xgas}).  This
temperature increase is likely to underestimate that within the shell
by a factor of a few, as noted above, and so our total estimated power
of expansion $\sim 4 \times 10^{44}$ ergs s$^{-1}$ is a conservative
estimate. If used instead of cavity power, \source\ agrees with the
correlation of \citet{cavagnolo}, once the large intrinsic scatter
of the correlation is taken into account.

Well-defined jet structure means \source\ offers a rare
possibility of comparing the time-averaged jet power with the
instantaneous power in the kpc-scale jet.  If we adopt a viewing angle
of $10^\circ$ and beaming with $\delta = 5.7$, which we argued are
likely from modelling the N hotspot
(Sec.~\ref{sec:discussionhotspots}) and are consistent with other
properties of the jet (Sec.~\ref{sec:discussionjet}), a calculation of
the kinetic energy flux \citep[see Appendix of][]{schwartz} for knot D
for leptons only and minimum energy finds $\sim 5 \times 10^{43}$ ergs
s$^{-1}$.  Assuming this is representative of the instantaneous power
of the jet, and doubling to take into account the counterjet, we find
a shortfall of a factor $\sim 4$ compared with the total time-averaged
power above.  A modest proton content or small departure from minimum
energy in the jet would bring the powers into agreement.

\section{Summary}
\label{sec:summary}

We have reported new results for the jet, hotspots, lobes and
environment of \source, a source of great interest not only for the
jet-cloud interaction region of its HIC, but also as a nearby
example of the population of radio galaxies expected to be most typical of
radio-mode feedback.

While the radio source is two-sided on large scales, with two bright
lobes and hotspots, the first high-resolution radio map of the source
traces a northern jet more than halfway to its terminal hotspot, but
fails convincingly to detect counter-jet emission.  The source could
thus be orientated at relatively small angle $\theta$ to the line of
sight, although $\theta < 10^\circ$ would be unusual for a radio
galaxy with prominent two-sided large-scale structure.

The jet is by no means straight.  It extends for $\sim 1.5$~kpc as a
continuation of the inner component seen in VLBI, and in the direction
of the HIC, but then deflects at least twice in a zig-zag path before
the final deflection that is observed adjacent to the HIC.  A new
analysis of \hst\ data finds optical emission from gas near the jet
not only at the HIC but also at the location of the first
deflection, although the ionization state of this latter gas is unknown.

Using new \chandra\ data we have concluded that the hot component of
gas ($kT \sim 0.3$ keV) associated with the HIC, and almost certainly
binding it together, should have dissipated through expansion in $\sim
6$~Myr.  The HIC is thus a relatively young feature as compared with
the likely overall lifetime of the radio source.  An unexpected result
is the anomalously low metallicity of the HIC X-ray gas --- something
not seen elsewhere in the larger-scale atmosphere.  Deeper X-ray
observations are required to test whether this is an artifact of a
complex temperature structure, or a physical result that will need
considering in the interpretation of forthcoming spatially-resolved
optical spectroscopy of the HIC.  In either case the close association
of jet deflection in the new radio data with the location of the HIC
supports a strong jet-gas interaction.

We detect optical and X-ray emission from extended lengths of radio
jet, as well as the knot close to the HIC.  We interpret this emission
as synchrotron radiation. An explanation in terms of beamed
inverse-Compton scattering of the CMB in a minimum-energy jet, as
claimed for many quasar jets, would make the angle to line of sight
uncomfortably small.  Cases of X-ray synchrotron emission over large
lengths of an FRII jet are relatively rare, and \source\ is one of the
best examples.  For the knot close to the HIC we deduce an energy-loss
time comparable to the light-crossing time, and so deduce that
particle acceleration is continuous within a small region of this
knot.

The two hotspots are particularly interesting.  Only the southern one
emits X-rays in the true `hotspot' associated with jet termination,
where bulk motion is sonic and the radio emission
peaks.  This region in both hotspots is detected in our
ground-based optical observations. We deduce that these emissions are
synchrotron, and if there is a common acceleration mechanism in the
two hotspots, acceleration of electrons to a maximum energy
of $\sim 10^{13}$~eV is suggested.  

However, whereas the hotspot X-ray emission on the
southern (counterjet) side is concentrated at jet termination, in the
north, X-rays are seen only where the jet approaches the hotspot, with
the X-ray and radio peaks offset by $\sim 600$~pc.  This X-ray emission
is difficult to understand unless the jet is oriented at small angle to
the line of sight, close to the lower limit expected for radio galaxies ($\theta
\sim 10^\circ$), and has a bulk relativistic Lorentz factor $\Gamma \sim 5.7$.  In that case the X-rays could arise from beamed inverse-Compton
scattering of CMB and hotspot synchrotron radiation.  The
interpretation would help to explain the phenomenon
being seen in the jet-side hotspot but not the
counterjet, and would allow a minimum-energy magnetic field that is
similar in the jet and the terminal hotspot, although the interpretation may be in
conflict with the X-ray spectral index.  Deeper X-ray observations and
high resolution optical measurements with HST would help
the interpretation of this important situation where we are separating emission
from the plasma entering the point of jet termination from the
termination point itself.

The large-scale properties of \source\ are of particular important in
the context of radio-mode feedback.  We are rewarded by the new
observations displaying multiple features to confirm and measure
supersonic expansion of the lobes into the external medium.  There are
clear depressions in X-ray intensity corresponding to the locations
of the radio lobes, indicating that the lobes evacuate cavities in its
$kT \sim 1$ keV atmosphere.  The X-ray spectra from the regions of the
lobes allow separation of components of inverse-Compton scattered CMB
emission.  This has allowed us to estimate the lobe magnetic field
strengths and lepton densities, independently of a minimum-energy
assumption.  As is typical for radio lobes, we find that the magnetic
field strengths ($\sim 1.0$ and $0.7$~nT, in the north and south,
respectively) are roughly one third the minimum-energy values.

The gas distribution is complex, and more complete modelling is the
subject of a further paper, but the internal lobe pressure exceeds
that in external regions we have modelled by a factor of 6, and
greater if relativistic protons add to the lobe's
internal energy.  This should be driving a shock of Mach number
between 2.2 and 3 into the external gas. We find that the
kinetic and thermal energy of shocked gas dominates the time-averaged
jet power, $\sim  4 \times 10^{44}$ ergs s$^{-1}$, which 
agrees with our estimate of the
instantaneous jet power for only modest proton content or
a small departure from minimum energy in the jet if
indeed $\theta \sim 10^\circ$ and $\Gamma \sim 5.7$.
We expected relatively strong shocks like this to be
common around expanding radio lobes, but they have proved somewhat
elusive, probably largely due to a selection bias towards observing
rich atmospheres, where the contrast between shocked and ambient
plasma is harder to discern and the ability to separate
inverse-Compton lobe emission is more difficult.  In \source, shocked
gas is apparent morphologically as an intensity enhancement, shaping a
truncated figure of eight around the lobes.  One region is
sufficiently bright to allow spectroscopy to a precision sufficient to
allow us to conclude that there is a chance less than one per cent
that it is as cool as ambient gas in regions we have investigated.
Deeper X-ray observations will allow us to extend such studies to more
regions of the atmosphere, and reach a more complete
understanding of the energetics of feedback in this representative
source.

\section*{Acknowledgments}

We thank staff at Narrabri for help and advice before and during the
\atca\ observations, and the STFC for travel support.  The Australia
Telescope is funded by the Commonwealth of Australia for operation as
a National Facility managed by CSIRO.  We are grateful to the CXC for
its support of \chandra.
We thank the anonymous referee for careful reading of the
manuscript and helpful comments.

\end{document}